\documentclass[twocolumn]{aa}  

\usepackage{graphicx}
\usepackage{txfonts}
\usepackage{longtable}
\begin{document}

\title{Inferring physical parameters in solar prominence threads}

\author{M. Montes-Solís
	\inst{1,2}
	\and
	I. Arregui \inst{1,2}\
}

\institute{Instituto de Astrofísica de Canarias, E-38205 La Laguna, Tenerife, Spain
	\and
	Departamento de Astrofísica, Universidad de La Laguna, E-38206 La Laguna, Tenerife, Spain\\
	\email{mmsolis@iac.es}
}

\date{Received ; accepted }

\abstract
{High resolution observations have permitted to resolve the solar prominences/filaments as sets of threads/fibrils. However, the values of the physical parameters of these threads and their structuring remain poorly constrained.}
{We use prominence seismology techniques to analyse transverse oscillations in threads through the comparison between magnetohydrodynamic (MHD) models and observations.}
{We apply Bayesian methods to obtain two different types of information. We first infer the marginal posterior distribution of physical parameters, such as the magnetic field strength or the length of the thread, when a totally filled tube, a partially filled tube, and three damping models (resonant absorption in the Alfvén continuum, resonant absorption in the slow continuum, and Cowling's diffusion) are considered as certain. Then, we compare the relative plausibility between alternative MHD models by computing the Bayes factors.}
{Well constrained probability density distributions can be obtained for the magnetic field strength, the length of the thread, the density contrast, and parameters associated to damping models. When comparing the damping models of resonant absorption in the Alfvén continuum, resonant absorption in the slow continuum and Cowling's diffusion due to partial ionisation of prominence plasma, the resonant absorption in the Alfvén continuum is the most plausible mechanism in explaining the existing observations. Relations between periods of fundamental and first overtone kink modes with values around 1 are better explained by expressions of the period ratio in the long thread approximation, while the rest of the values are 
more probable in the short thread limit for the period ratio.}
{Our results show that Bayesian analysis offers valuable methods for performing parameter inference and model comparison in the context of prominence seismology.}


\maketitle
  %
  
\section{Introduction}
Understanding the physical plasma conditions, dynamics, and energetics of solar prominences is a challenge. High resolution imaging observations in H$_{\alpha}$ with e.g., the Swedish Solar Telescope (SST) and Hinode have enabled us to resolve the fine structures forming the prominences bodies. These structures consist of fine threads not thicker than $\sim$ 0.2-0.3 arc sec ($\sim$ 145 - 218 km) \citep{Lin2004,Lin2005,Okamoto2007,Lin2008}. Characterising the physical properties and dynamics of these fine threads is key to understand the whole solar prominences. 

To complement direct observations, indirect inference methods offer valuable information. Ample observational evidence exists about the presence of waves in prominences \citep{Oliver2002,Arregui2018_review}. These observed waves are classified in large and small amplitude oscillations and imply both transverse or longitudinal motions with respect to the structures \citep[see e.g.][]{Yi1991,Yi1991b,Terradas2002,Foullon2004,Berger2008,Tripathi2009}. Focusing on small amplitude oscillations, they display a wide range of periods and typical velocity amplitudes consistent with the time-scales of transverse motions in threads reported by \citet{Lin2007,Lin2009}. Mass flows have also been detected, with velocities in the range 15 -- 46 km s$^{-1}$ \citep{Zirker1998,Lin2005,Okamoto2007,Ning2009}. These plasma flows can possibly affect wave propagation properties.

Several theoretical models with different levels of complexity have been proposed to explain prominence thread oscillations. Simple fibril models under the zero plasma-$\beta$ approximation in Cartesian geometry \citep{Joarder1997,Diaz2001} laid the foundations of thread oscillation studies to more complex and realistic models in cylindrical geometry focused on fast MHD wave properties \citep{Diaz2002,Dymova2005}. Posterior studies by \citet{Soler2011} considered the influence of mass flows on the different oscillating modes. 

Damping of transverse waves in threads is a commonly observed phenomenon with typical ratios of the damping time to the period in the range 1 to 10 \citep{Arregui2011a}. Diverse mechanisms have been considered to explain the damping process in threads. Non-ideal effects were considered in uniform media with and without flows \citep{Carbonell2004,Terradas2005, Carbonell2006,Carbonell2009}, in stratified media with and without flows \citep{Soler2007,Soler2008, Soler2009a} and with partial ionisation effects \citep{Forteza2007,Forteza2008}. Among partial ionisation effects, only Cowling's diffusion seems to be efficient enough to produce strong damping compatible with observations \citep{Soler2009b,Soler2009c}. Wave leakage has been also considered as damping mechanism \citep{Oord1992,Schutgens1997,Schutgens1997b,Oord1998} but some issues such as different damping rates for vertical and horizontal oscillations still remain unclear \citep{Schutgens1999,McLaughlin2006,McDougall2007}. Resonant absorption of kink waves in the Alfvén continuum, first proposed by \citet{Arregui2008} in this context of prominence threads, is another candidate to explain this phenomenon with damping time-scales compatible with those observed. In a subsequent work by \citet{Soler2009d}, damping due to resonant absorption in the slow continuum was further considered, but for typical prominence parameters, the resulting damping times seems to be too long in comparison to those observed. Additional information on damping mechanisms for prominence oscillations can be found in \cite{Arregui2011_review} and \cite{Arregui2018_review}.

Over the years, improvements in observations and theory have enabled us to obtain more precise measurements of wave properties and a better understanding of theoretical prominence thread oscillations. Improving the accuracy of seismology inversions however requires the adoption of new techniques for the comparison between observations and theory and the solution to the ensuring inverse problems. Recent developments in coronal loop seismology have considered the use of Bayesian inversion techniques \citep{Pascoe2017,Pascoe2018,Arregui2018} which have been successful in the task of obtaining information about physical parameters from observations (inference) and in comparing the relative performance of alternative models to explain observed data (model comparison). Motivated by the success of Bayesian methods in coronal seismology, we aim to apply them to solar prominence seismology to infer physical parameters of threads and to compare between different theoretical models. 

The layout of the paper is as follows. The Bayesian methodology used in our study is explained in Section~\ref{sec:bayes}.  Then, in Section~\ref{sec:results} our results on parameter inference and model comparison are presented. Our summary and conclusions can be found in Section~\ref{sec:conclusions}. 
  
\section{Methodology: Bayesian statistics}\label{sec:bayes}
  
Bayesian methods permit to confront theoretical models and observations in three different levels. At a first level, we can perform inference of model parameters assuming a particular considered model as certain. Then, at a second level, we can compare the plausibility of alternative models in explaining observed data. These two levels have already been applied in coronal loops by \cite{Yo2017}. A third level, not considered here, would consist of model averaging \citep[see e.g.,][]{Arregui2015}.

The foundational principle for all three inference levels is the Bayes' theorem defined as 
\begin{equation}
\label{eq1}
p(\boldsymbol{\theta}|M,d)=\frac{p(\boldsymbol{\theta}|M) p(d|M,\boldsymbol{\theta)}}{p(d|M)}, 
\end{equation}
which establishes the posterior distribution $p(\boldsymbol{\theta}|M,d)$ of a set of parameters, $\boldsymbol{\theta}$, assuming one particular model $M$ as certain, and conditional on observed data, $d$. This probability density distribution gives a simple relation between the prior probability, $p(\boldsymbol{\theta}|M)$, before considering the data; and the likelihood function, $p(d|M,\boldsymbol{\theta})$. The denominator in Eq.~(\ref{eq1}) is a normalisation factor, the so-called marginal likelihood of the data given a model $M$. It can be computed by performing the integral
\begin{equation}
\label{eq2}
p(d|M)=\int_{\boldsymbol{\theta}}{p(\boldsymbol{\theta}|M)p(d|M,\boldsymbol{\theta})d\boldsymbol{\theta}}
\end{equation} 
over the full parameter space so the probability of the data given the model is obtained.

When model $M$ is represented in terms of $n$ parameters, the inference of information on each specific parameter $\theta_{i}$ can be gathered by computation of the marginal posterior, integrating the full posterior with respect to the rest of $n$ model parameters as  
\begin{equation}
\label{eq3}
p(\theta_{i}|M,d)=\int{p(\boldsymbol{\theta}|M,d)d\theta_1...d\theta_{i-1}d\theta_{i+1}...d\theta_{n}}. 
\end{equation}   

Moving to model comparison, Bayesian methods offer tools to quantify the plausibility of different mechanisms in explaining the observed data in two ways.  Marginal likelihoods corresponding to alternative models can be computed, using Eq.~(\ref{eq2}), and directly compared. On the other hand, if there is no a priori preference for any of the considered models, a one-to-one comparison can be performed using the so-called Bayes factor \citep[see e.g.,][]{Kass1995} defined as
\begin{equation}
\label{eq4}
BF_{\rm kj}=\frac{p(d|M_{\rm k})}{p(d|M_{\rm j})} \ ;\ k\neq j.
\end{equation} 
By following the evidence classification criteria by \citet{Kass1995}, see Table \ref{tab1}; the level of evidence for one model against the alternative is evaluated. 

\begin{table}[h]
	\caption{\citet{Kass1995} evidence classification.\label{tab1}}
	\centering
	\begin{tabular}{c c}
		\hline\hline
		$2\ ln BF_{\rm kj}$ & Evidence\\
		\hline
		0-2 & Not Worth more than a bare Mention (NWM)\\
		2-6 & Positive Evidence (PE)\\
		6-10 & Strong Evidence (SE)\\
		$>$10 & Very Strong Evidence (VSE)\\
		\hline
	\end{tabular}
\end{table}

Any Bayesian inference, at any level, will depend on the prior information one accepts as plausible and on the likelihood function. These two pieces of information are adopted and lead to the posteriors, which are computed. In this study, we have adopted independent priors for model parameters, so that the global prior is given by the product of the individual priors associated with each parameter. We have considered two types of individual prior. To express our prior belief that each parameter lies on a given plausible range, with all values being equally probable a priori, a uniform prior of the form 
\begin{equation}
\label{eq5}
p(\theta_{\rm i})=\frac{1}{\theta_{\rm i,max}-\theta_{\rm  i,min}},
\end{equation}
has been adopted. On the other hand, to take into account some more specific prior information on a particular parameter of interest, a Gaussian prior of the form
\begin{equation}
\label{eq6}
p(\theta_{\rm i})=\frac{1}{\sqrt{2\pi}\sigma_{\rm \theta_{\rm i}}}\exp{\left[{-\frac{\left(\theta_{\rm i}-\mu_{\rm \theta_{\rm i}}\right)^2}{2\sigma^2_{\rm \theta_{\rm i}}}}\right]}\,
\end{equation}
has been considered, where $\mu_{\rm \theta}$ and $\sigma_{\rm \theta}$ represent the estimated value of the parameter and its uncertainty, respectively. As likelihood function, Gaussian profiles have been applied according to normal errors assumption.

The computation of posteriors and marginal likelihoods further requires to solve integrals in the parameter space. When the problem at hand is low-dimensional, direct numerical integration will be possible, with Monte Carlo (MC) methods as a feasible alternative in low-dimensional but also applicable in high-dimensional spaces. As in \citet{Yo2017} we have followed both approaches by making sure both give identical results.

\section{Results}\label{sec:results}

In this paper, Bayesian methods are applied to a number of theoretical models and observations of transversally oscillating prominence threads. We first do inference of some physical parameters such as the magnetic field strength or the length of the thread. Then, we compare alternative models for the damping of oscillations. Finally, the relation between the periods of the fundamental and first overtone kink modes is considered to compare short and long thread approximations.

\subsection{Inference}
\subsubsection{Magnetic field strength}
We first assume that threads can be modelled as homogeneous thin flux tubes with an internal prominence density, $\rho_{\rm p}$, in a coronal environment with density $\rho_{\rm c}$. Under this assumption, the phase speed of fast kink modes in threads can be simply expressed as $v_{\rm ph}=v_{\rm Ai}\sqrt{2\zeta/(1+\zeta)}$, which depends on the internal Alfvén velocity, $v_{\rm Ai}$, and the density contrast, $\zeta=\rho_{\rm p}/\rho_{\rm c}$ \citep{Spruit1982,Edwin1983}. 

Expanding the Alfvén speed as a function of the magnetic field strength, $B$, and the internal density in the thread, the theoretical equation for the phase speed can be written in the following form 
\begin{equation}
\label{eq7}
v_{\rm ph}=\sqrt{\frac{2}{\mu_{0} \rho_{\rm p}}}B, 
\end{equation}
which is a good approximation for sufficiently large density contrasts typical of these structures (see panel a of figure 3.8 by \citealt{Soler2010b}).

Observationally, the phase speed of transverse kink waves in prominence threads can be estimated \citep{Lin2009}. By assuming that the phase speed estimated from observations is equal to the kink speed, $v_{\rm ph}\approx c_{\rm k}$, given in Eq.~(\ref{eq7}) and taking this as the observable data, $d=v_{\rm ph}$, the inference of the unknown parameters $\boldsymbol {\theta}=\{ \rho_{\rm p},B\}$ can be attempted. 

\begin{figure}[!h]
	\centering
	\includegraphics[scale=0.40]{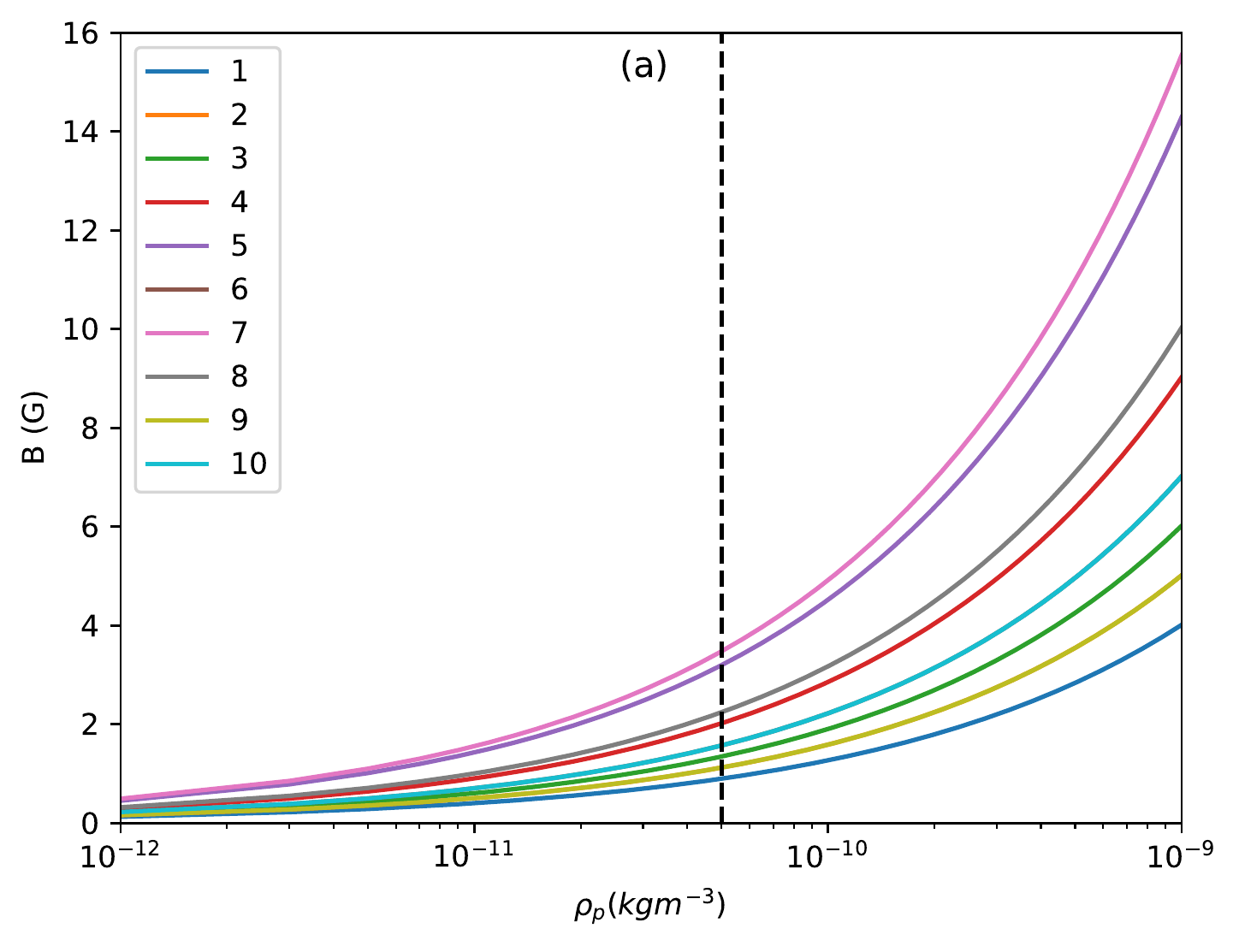}\\
	\includegraphics[scale=0.40]{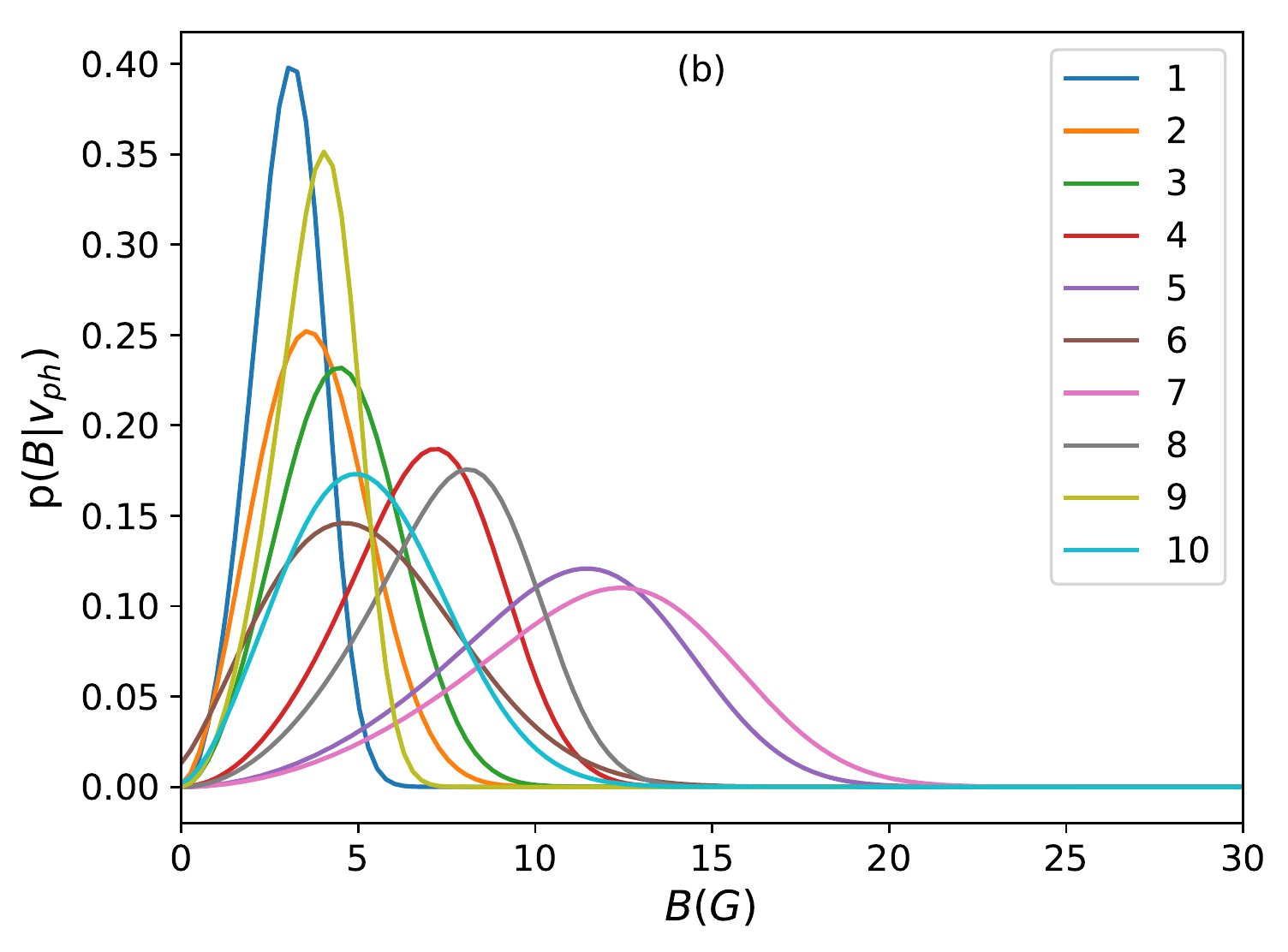}\\
	\includegraphics[scale=0.45]{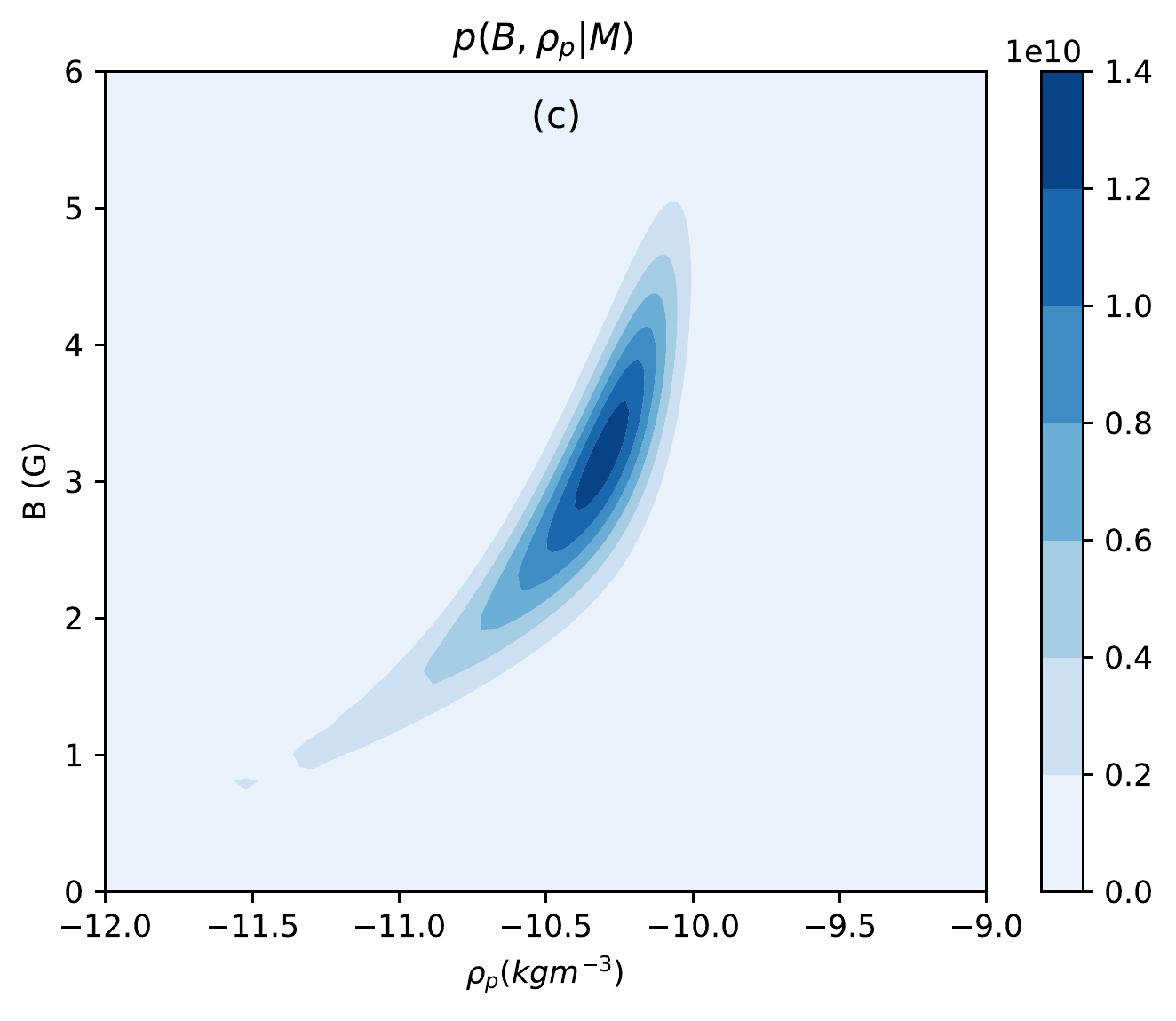}\\
	\includegraphics[scale=0.40]{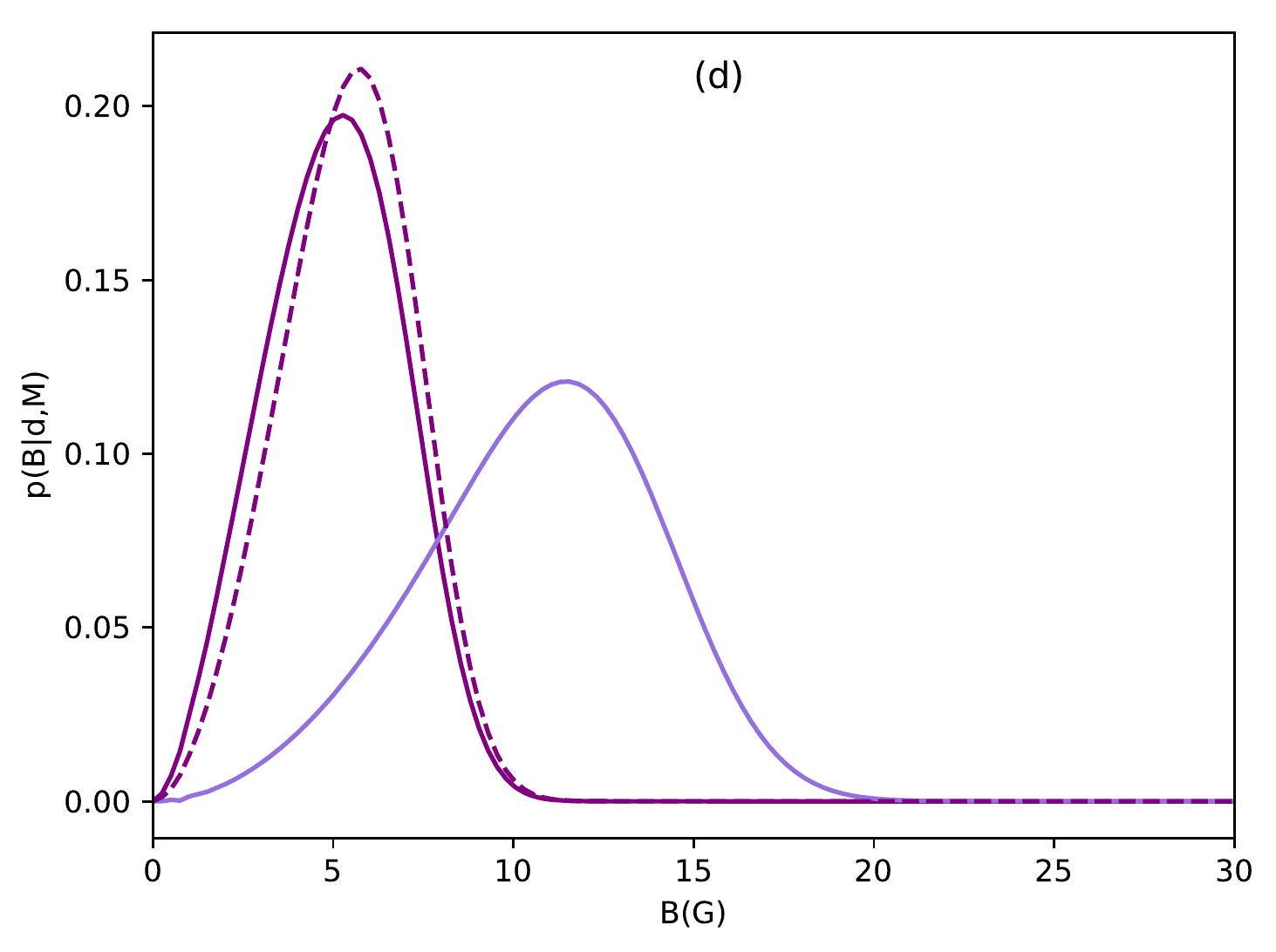}\\
	\caption{(a) Curves obtained for each observed thread by \citet{Lin2009} in the forward problem using Eq.~(\ref{eq7}). (b) Posterior distributions of magnetic field strength ($B$) for each considered thread obtained with Bayesian methods. Each number and colour in (a) and (b) correspond to each observed thread according to legends and Table~\ref{tab2}. (c) Global posterior computed for the fifth thread, considering a Gaussian prior of the internal density ($\rho_{\rm i}$) centred in a value equal to $\rho_{\rm p}=5\times 10^{-11}$ kg m$^{-3}$ and given an uncertainty of 50 \%. (d) Comparison of posterior distributions of the magnetic field strength in totally filled tube (purple line), partially filled tube (magenta line), and partially filled tube considering a Gaussian prior for the proportion of thread length centred in $L_{\rm p}/L=0.5$ with 50\% of uncertainty (dashed magenta line). \label{f1}}
\end{figure}

In their analysis, \citet{Lin2009} took a fixed value of $\rho_{\rm p}=5\times 10^{-11}$ kg m$^{-3}$ and solved Eq.~(\ref{eq7}) for the magnetic field strength in a forward analysis. Their results for the 10 analysed threads are shown in the third column of Table~\ref{tab2}. The assumption of a particular value of $\rho_{\rm p}$ enables in principle to estimate the magnetic field strength. The main drawback of this procedure is that the density of prominence plasmas is highly uncertain and difficult to estimate with some accuracy. Repeating the procedure for values of density in an extended plausible range leads to a variability in the inferred magnetic field strength, see Fig.~\ref{f1}a. 

In this paper, we have performed the Bayesian inference of the magnetic field strength by assuming uniform priors in Eq.~(\ref{eq5}) for both parameters. Assuming plausible ranges for $B\in[0.01,50]$ G and $\rho_{\rm p}\in[10^{-12},10^{-9}]$ kg m$^{-3}$, theoretical model predictions cover the range of phase speeds $v_{\rm ph}\in(0-6000)$ kms$^{-1}$. 

\begin{figure}[h]
	\centering
	\includegraphics[scale=0.43]{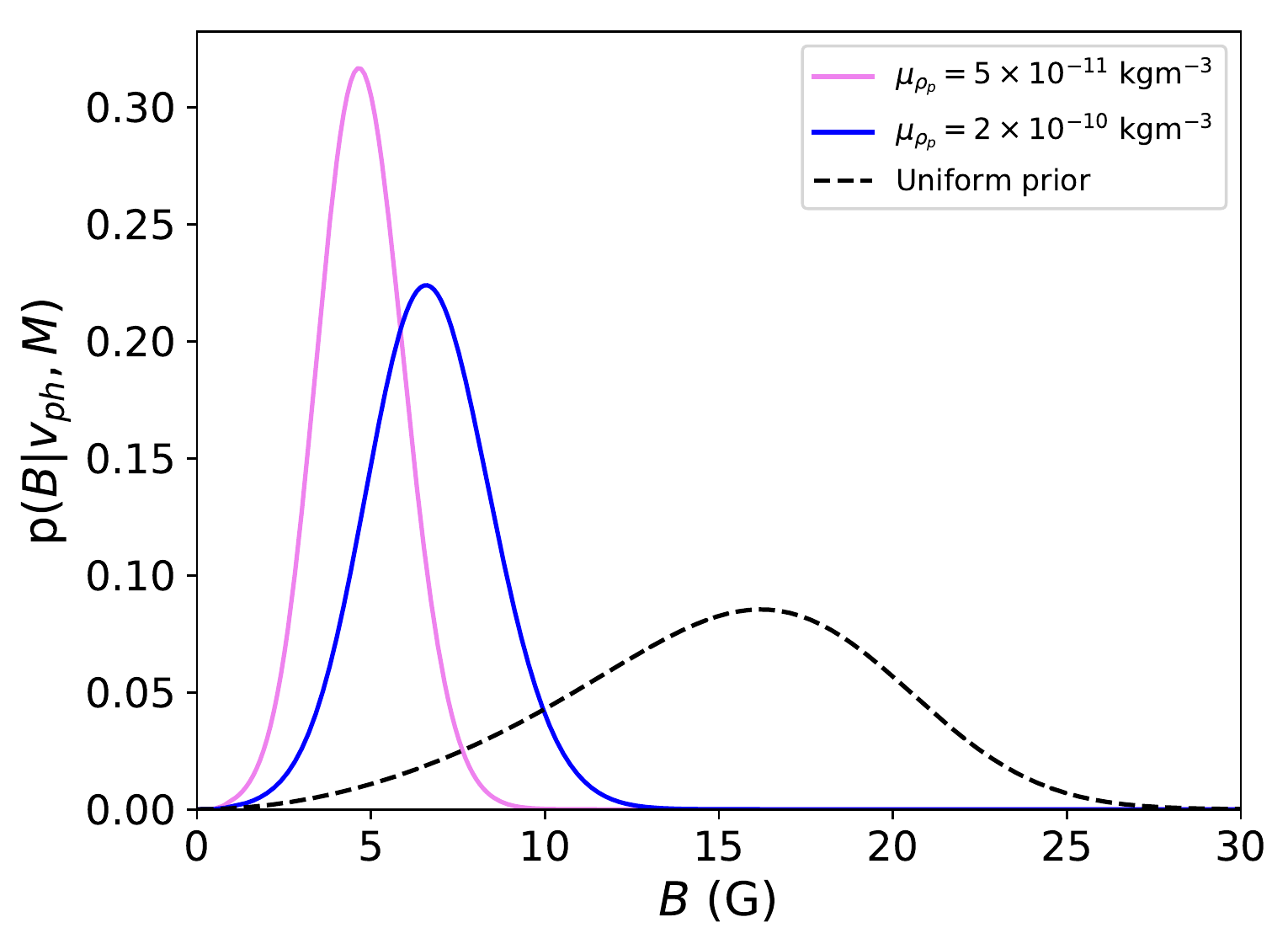}
	\caption{Posterior distributions of the magnetic field strength (B) considering a Gaussian prior of the internal density ($\rho_{\rm i}$) centred in a value equal to $\rho_{\rm p}=5\times 10^{-11}$ kg m$^{-3}$ (pink), a Gaussian prior centred in a value equal to $\rho_{\rm p}=2\times 10^{-10}$ kg m$^{-3}$ (blue), and a uniform prior (black dashed line). Uncertainties in Gaussian priors are considered to be of 50 \%. The totally filled tube and the plausible ranges for the parameters of Fig.~\ref{f1} have been assumed.\label{f2}}
\end{figure}

Figure~\ref{f1}b shows the Bayesian inference result obtained for each thread analysed by \citet{Lin2009} in terms of marginal posterior distributions for the magnetic field strength. The median and errors at the 68\% credible interval are given in Table~\ref{tab2}. For all threads, the marginal posteriors can be properly inferred. Figure~\ref{f1}b shows that the distributions spread over a range of values ranging from below 1 up to 20 G. When we compare to the classic result in Fig.~\ref{f1}a, the difference now is that the values of magnetic field strength that spread over a given range because of the variability in density, have now with the Bayesian analysis a different plausibility. It is worth noting that the distributions for different threads belonging to the same prominence display rather different maximum a posteriori estimates. This is indicative of the highly inhomogeneous nature of the magnetic field strength at small scales. Note that the transverse non-uniformity of the field strength would need to be counteracted by gas pressure forces to provide the constant total pressure in the transverse direction, an ingredient that has not been considered in the model under consideration. If we compare our results using uniform priors and a range of density values (column $B_{\rm u}$ in Table~\ref{tab2}) with those by \citet{Lin2009} using a fixed prominence density (column $B_{\rm Lin}$ in Table~\ref{tab2}) we appreciate significant differences. The results by \cite{Lin2009} seem to be more precise and ours more uncertain. This is because \cite{Lin2009} remove all the uncertainty on the prominence density. Instead of fixing a value of the density as \cite{Lin2009} do, Bayesian methods enable us to mimic this by considering a Gaussian prior centred in $\rho_{\rm p}=5\times 10^{-11}$ kg m$^{-3}$ and with some uncertainty. An example result is shown in Fig.~\ref{f1}c where the joint posterior for the magnetic field strength and the density for the thread number five is plotted. Now, comparing the column $B_{\rm Lin}$ with the column $B_{\rm G}$ corresponding to calculations with Gaussian priors in Table~\ref{tab2}, we see that we obtain similar results, with the advantage that the Bayesian results have the uncertainty correctly propagated. If we compare the result for thread number five using uniform prior (purple line in Fig.~\ref{f1}b or its summary value in Table~\ref{tab2}), with the one obtained using the Gaussian prior, we see that a better constrained magnetic field strength is inferred for this latter case. The accuracy in the magnetic field strength inference thus depends strongly on the precision in observationally estimated prominence densities. In the absence of information on the internal density, this parameter cannot be properly inferred, as has been noted by \cite{Arregui2018b} in the context of coronal loop oscillations.
Figure~\ref{f2} shows posterior distributions of the magnetic field strength, for the fifth thread, computed using different priors for the density. Besides the uniform prior, two Gaussian priors centred at two density values are used. The results for each case clearly differ.

The cool and dense thread in prominences does not occupy the entire magnetic flux tube. If we now consider the thread with length $L_{\rm p}$ as part of a longer flux tube of total length $L$, the period can be approximated by $P\approx\pi\sqrt{L_{\rm p}(L-L_{\rm p})}/\sqrt{2}v_{\rm Ai}$ \citep{Soler2010}. This period can be easily transformed to phase velocity ($v_{ph}=2L/P$) with the following form
\begin{equation}
v^{par}_{\rm ph}=\frac{2}{\pi\sqrt{\frac{L_{\rm p}}{L}\left(1-\frac{L_{\rm p}}{L}\right)}}v^{tot}_{\rm ph},
\end{equation}
where we have denoted $v^{tot}_{\rm ph}$ for the phase velocity in Eq.~(\ref{eq7}) for the totally filled tube and $v^{par}_{\rm ph}$ for a partially filled tube. In this model, the space of parameters becomes larger with the ratio between the length of the thread and the total length, $L_{\rm p}/L$, as an additional parameter. Under these conditions, we inferred the magnetic field strength for each thread analysed by \citet{Lin2009} to quantify the differences with the previous results. Figure~\ref{f1}d shows an example of comparison between marginal posteriors for the magnetic field strength in a totally filled tube (purple line) and in a partially filled tube (magenta line) for the thread number five. Uniform priors have been considered for all parameters with the length of the thread to the total length in the plausible range $L_{\rm p}/L\in(0,1)$. The posterior distribution for the partially filled tube shows more constrained values of the magnetic field strength than the posterior corresponding to the totally filled tube (see median values at the sixth column of Table~\ref{tab2}). Additionally, it peaks at a smaller value of this parameter. Figure~\ref{f1}d further includes a dashed magenta line resulting from considering a Gaussian prior for the new parameter centred at $L_{\rm p}/L=0.5$ with an uncertainty of 50\%. This solution is very similar to the one that considered a uniform prior over the full range of $L_p/L$.
\begin{table}[h]
	\caption{Summary of results from the analysis of threads observed by \citet{Lin2009}. The columns contain the thread number (\#), the phase velocity ($v_{\rm ph}$), the magnetic field derived by \cite{Lin2009} ($B_{\rm Lin}$), and the median and errors at the 68\% credible interval for the magnetic field strength computed using uniform priors ($B_{\rm u}$), Gaussian priors ($B_{\rm G}$), and for a partially filled tube using uniform priors ($B^{\rm par}_{\rm u}$). \label{tab2}}
	\centering
	\begin{tabular}{cccccccc}
		\hline \hline 
		Thread &  $v_{\rm ph}$ & $B_{\rm Lin}$ & $B_{\rm u}$& $B_{\rm G}$ &$B^{\rm par}_{\rm u}$\\
		\#& (km s$^{-1})$ &  (G) & (G) & (G) & (G)\\
		\hline 
		1 & $16 \pm 3$ & $0.9\pm 0.3 $ &$3^{+1}_{-1}$ &$1.0^{+0.3}_{-0.2}$ &$2\pm1$\\ \\
		2 & $20 \pm 6$ & $1.1\pm 0.5 $ &$4^{+2}_{-2}$ &$1.2^{+0.5}_{-0.4}$ &$2\pm 1$\\ \\
		3 & $24 \pm 6$ & $1.3\pm 0.5 $ &$5^{+2}_{-2}$ &$1.5\pm 0.5$ &$2\pm1$\\ \\
		4 & $36 \pm 6$ & $2.0\pm 0.4 $ &$7^{+2}_{-2}$ &$2.2\pm 0.6$ &$4\pm 1$\\ \\
		5 & $57 \pm 9$ & $3.2\pm 0.7 $ &$11^{+4}_{-3}$ &$3.4\pm 0.9$ &$5\pm 2$\\ \\
		6 & $28 \pm 12$ & $1.6\pm 0.9 $ &$5^{+3}_{-3}$ &$1.7^{+0.8}_{-0.7}$ &$3^{+2}_{-1}$\\ \\
		7 & $62 \pm 10$ & $3.5\pm 0.8 $ &$12^{+4}_{-4}$ &$4\pm 1$ &$6\pm 2$\\ \\
		8 & $40 \pm 6$ & $2.3\pm 0.8 $ &$8^{+3}_{-2}$ &$2.4\pm 0.6$ &$4\pm 1$\\ \\
		9 & $20 \pm 3$ & $1.1\pm 0.2 $ &$4^{+1}_{-1}$ &$1.3\pm 0.3$ &$2\pm 1$\\ \\
		10 & $28 \pm 9$ & $1.6\pm 0.7 $ &$5^{+2}_{-3}$ &$1.7^{+0.7}_{-0.6}$ &$3^{+2}_{-1}$\\ 
		\hline\hline
	\end{tabular}
\end{table}

\subsubsection{Damping model parameters}\label{sec:damping}

Damping of transverse oscillations is a commonly observed phenomenon in threads, see e.g. \citet{Ning2009}. A number of theoretical models have been put forward to explain the damping process in prominences \citep{Arregui2011a}. These two ingredients of damping, observations and theoretical models, are used to perform seismology and infer conditions in these structures \citep[see][for a review]{Arregui2018_review}. In this work, we consider three damping mechanisms, namely resonant absorption in the Alfvén continuum \citep{Arregui2008}; resonant absorption in the slow continuum \citep{Soler2009d}; and Cowling's diffusion in a partially ionised plasma \citep{Soler2009b,Soler2009c} to extract information of physical features in prominence threads.

The first considered damping model is resonant absorption in the Alfvén continuum. This is a widely invoked mechanism in the context of coronal loop oscillations \citep{Goossens2002} and was suggested in the context of prominence plasmas by \citet{Arregui2008}. The mechanism  transfers global kink mode energy to small scale azimuthal motions at the boundary of the flux tube that separates the prominence and coronal plasmas. Under the thin tube and thin boundary approximations and considering large density contrast ratios, $\rho_{\rm p}>>\rho_{\rm c}$, a simple expression relates the damping ratio and the transverse inhomogeneity length scale,
\begin{equation}
\label{eq8}
\centering
\frac{\tau_{\rm d}}{P}=\frac{2}{\pi}\frac{R}{l}.
\end{equation}
In this equation, $\tau_{\rm d}$ is the damping time, $P$ the oscillation period, and $l/R$ the inhomogeneity length scale at the tube boundary in units of the tube radius R. The factor $2/\pi$ arises because we are considering a sinusoidal variation of density at that tube boundary.

\begin{figure}[!h]
	\centering
	\includegraphics[scale=0.45]{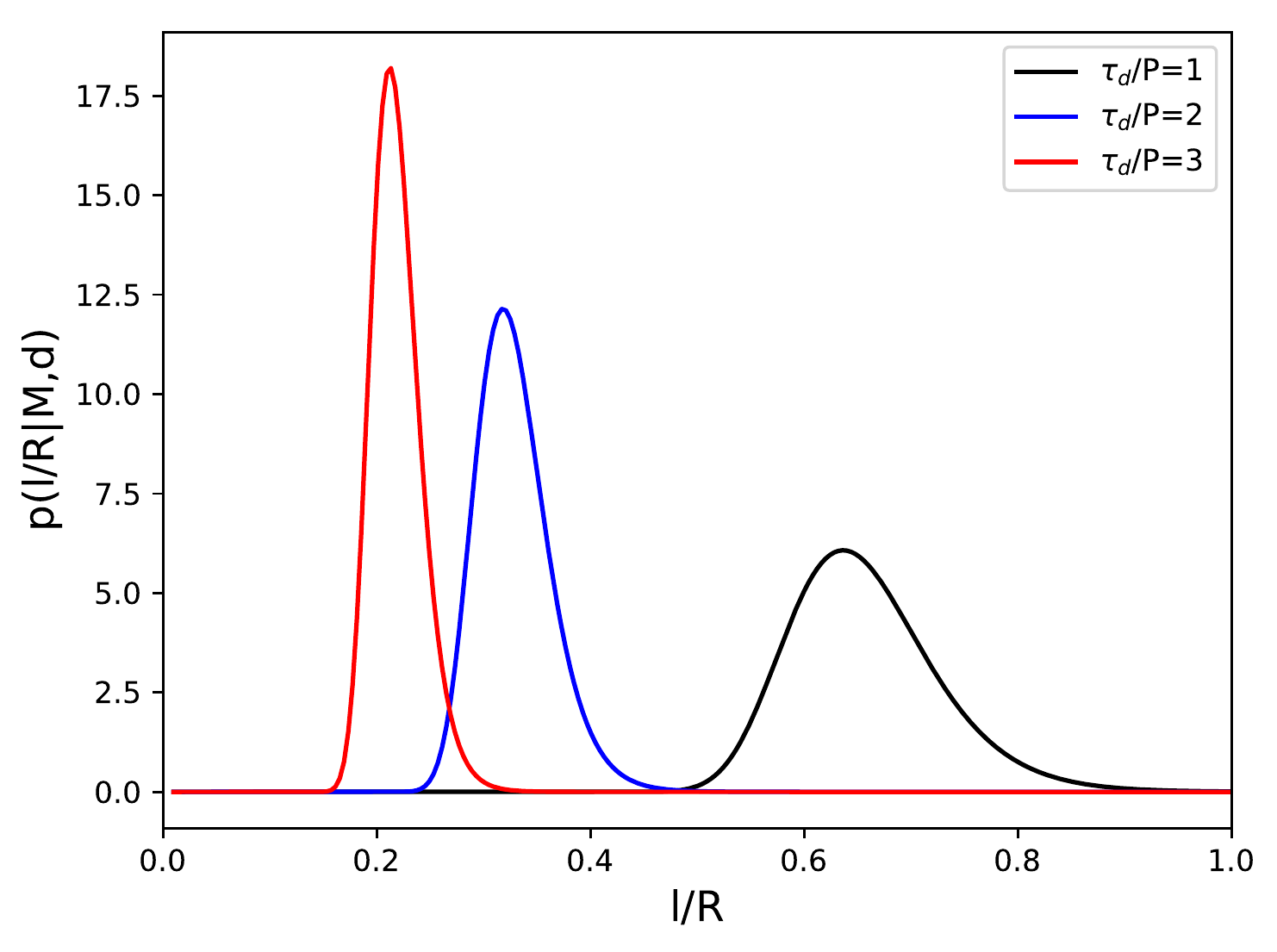}	
	\caption{Posterior distributions of the thickness of non-uniform layer ($l/R$) for resonant absorption in the Alfvén continuum. Three different damping ratios ($\tau_{\rm d}/P$) with an associated uncertainty of 10\% are considered. \label{f3}}	
\end{figure}

The use of observational values of the damping ratios, $d=\tau_{\rm d}/P$, and the theoretical expression in Eq.~(\ref{eq8}), permit to infer the damping model parameter for resonant absorption in the Alfvén continuum, $\boldsymbol {\theta}=\{ l/R \}$. The possible values that $l/R$ can take are restricted by the model itself with $l/R=0$ for a tube with a jump on density between $\rho_{\rm p}$ and $\rho_{\rm c}$ and $l/R=2$ for a fully non-uniform continuous variation of density. This leads to theoretically predicted damping ratios in the range $\tau_{\rm d}/P\approx[0.3,{\boldsymbol \infty }]$, compatible with observed values. 

Figure~\ref{f3} shows resulting posterior distributions for the damping model parameter when a uniform prior and three different values of the damping ratio with a given uncertainty are considered. All distributions can be inferred with short values of $l/R$ being more plausible for larger damping ratio values. These results are akin to those obtained by \citet{Yo2017} for the  transverse inhomogeneity length scales in coronal loop oscillations damped by resonant absorption.


The next considered damping model is resonant absorption in the slow continuum. In this case, the damping is produced by an energy transfer from the global kink mode to slow mode oscillations at the boundary of the magnetic flux tube. Relaxing the zero-$\beta$ approximation, the existence of slow magneto-acoustic waves cannot be despised since plasma in threads has chromospheric properties. This mechanism was first considered by \citet{Soler2009d} in the context of prominence threads. They derived an analytical expression for the damping ratio of the form
\begin{equation}
\label{eq9}
\frac{\tau_{\rm d}}{P}=\frac{2}{\pi}\frac{R}{l}\left (\frac{k_{\rm z}R}{1+\frac{2}{\gamma \beta}}\right)^{-2},
\end{equation}
with $k_{\rm z}R$ the longitudinal wave-number normalised to the radius of the tube, $\gamma=5/3$ the adiabatic constant of a mono-atomic gas, and $\beta$ the plasma-$\beta$ parameter. 

\begin{figure}[!h]
	\centering
	\includegraphics[scale=0.4]{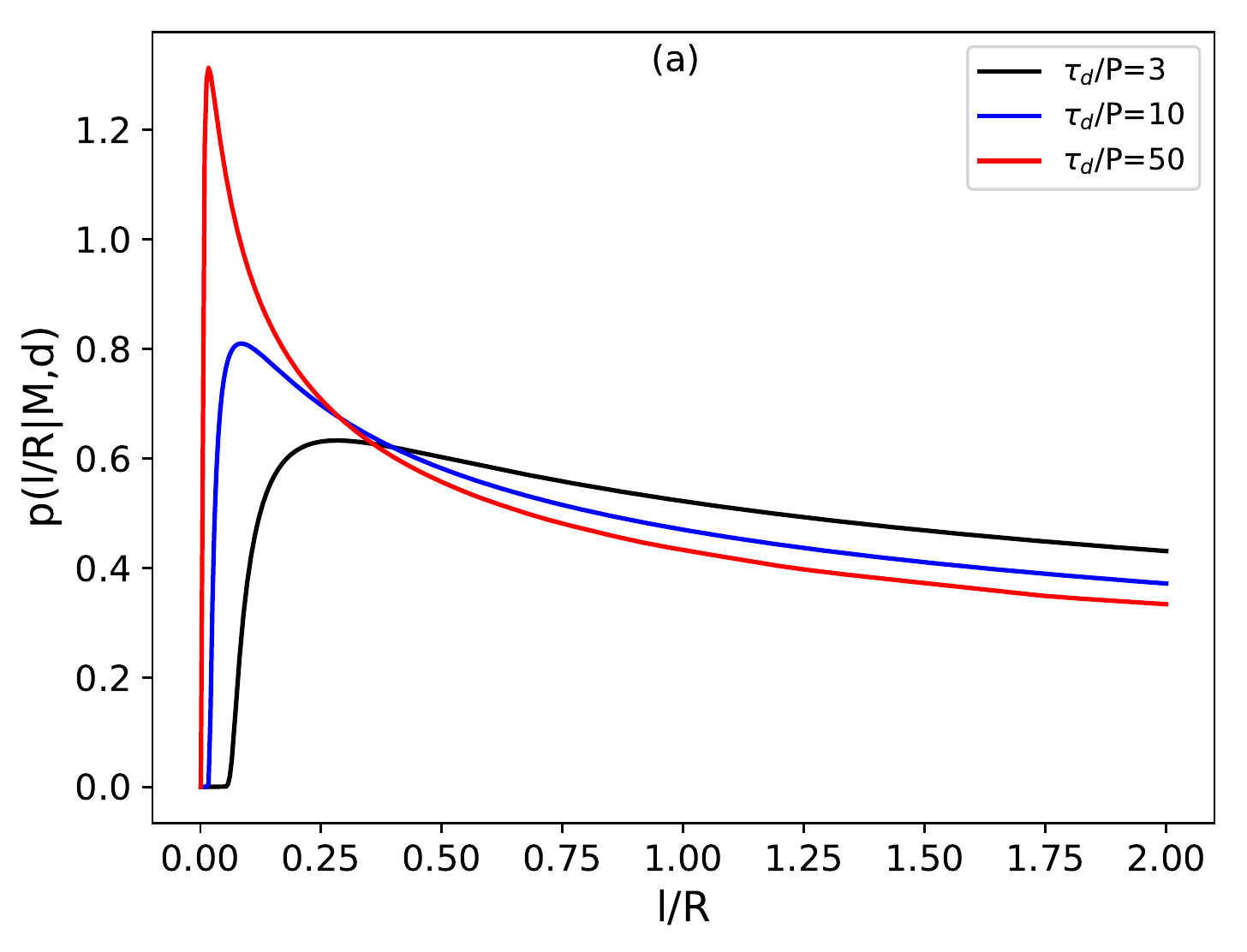}
	\includegraphics[scale=0.4]{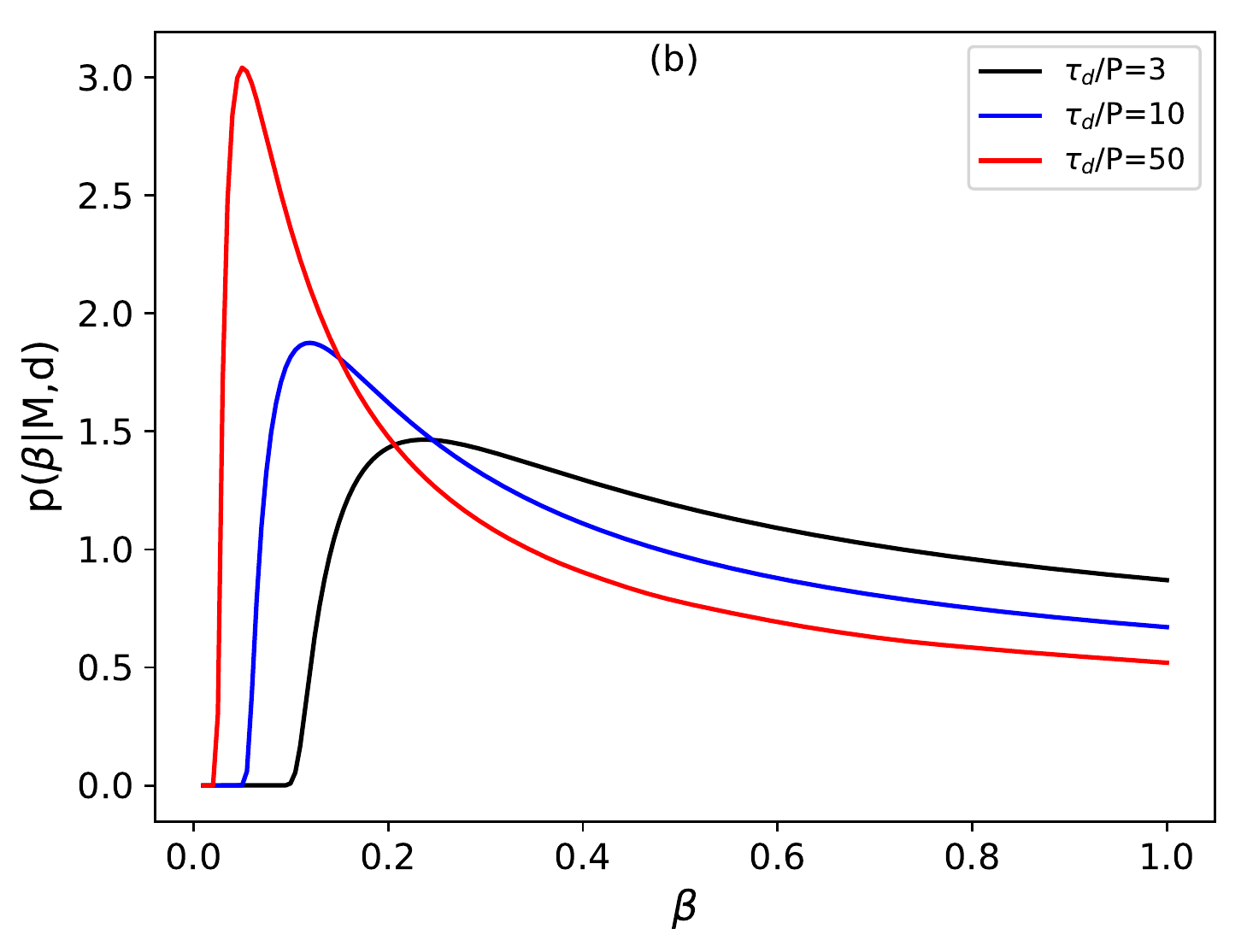}
	\includegraphics[scale=0.4]{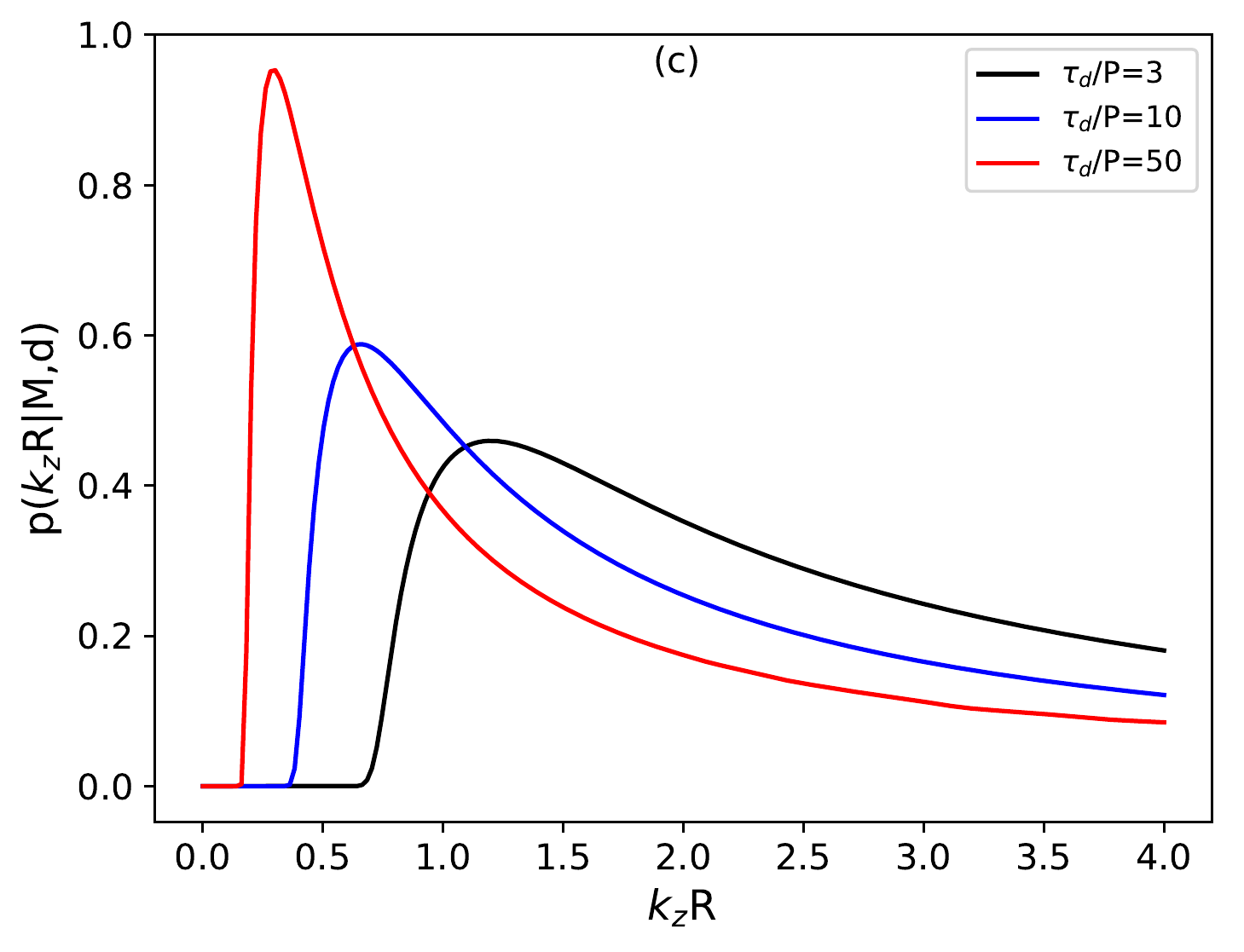}	
	\caption{Marginal posterior distributions for the width of non-uniform layer ($l/R$), plasma-$\beta$ parameter ($\beta$), and wave-number ($k_{\rm z}R$) for resonant absorption in the slow continuum. Different damping ratios ($\tau_{\rm d}/P$) with an associated uncertainty of 10\% are considered. \label{f4}}	
\end{figure}

Taking the damping ratio as observable, $d=\tau_{\rm d}/P$, and Eq.~(\ref{eq9}), we can attempt to infer the three parameters associated to damping by resonant absorption in the slow continuum, $\boldsymbol {\theta}=\{ l/R, \beta, k_{\rm z}R \}$. Plausible ranges of the model parameters $l/R\in[0.01,2]$, $\beta\in[0.01,1]$, and $k_{\rm z}R\in[10^{-3},0.1]$ lead to theoretically predicted damping ratios in the range $\tau_{\rm d}/P\thicksim[154,10^{13}]$. Figure~\ref{f4} shows the resulting marginal posteriors associated to the three model parameters for resonant absorption in the slow continuum, with uniform prior assumptions for all parameters. All distributions can be inferred with the same tendency to peak in larger parameter values for smaller damping ratio values. At the left panel, posterior distributions associated to the thickness of non-uniform layer peak at a similar value of $l/R$, regardless of the value of $\tau_{\rm d}/P$, a value that is similar to the $l/R$ value for resonant absorption in the Alfvén continuum. At the middle panel, posterior distributions for the plasma-$\beta$ parameter peak at small values indicating that the zero plasma-$\beta$ approximation in prominence threads seems a good approximation. At the right panel, posteriors have been also inferred for $k_{\rm z}R$ for the three considered damping ratios, but the range of possible wave-number values in threads had to be extended until $k_{\rm z}R=4$, otherwise this mechanism is not able to explain those damping ratios. Much larger damping ratios need be considered to have a proper inference with the typical reduced wave-numbers. 

We could ask ourselves if results would change considering a more realistic adiabatic constant. Repeating the inference process with the value derived by \citet{Doorsselaere2011} ($\gamma=1.10\pm0.02$) to define a Gaussian prior for the adiabatic index $\gamma$, we found that the posterior distributions of the model parameters do not differ significantly.

Finally, we consider the mechanism of Cowling's diffusion in partially ionised plasmas. This process consists on a magnetic diffusion in the perpendicular direction to the magnetic field lines due to ion-neutral collisions. Considering the analysis by \citet{Soler2009b,Soler2009c}, the theoretical expression for the damping ratio can be expressed as  
\begin{equation}
\label{eq10}
\frac{\tau_{\rm d}}{P}=\frac{\sqrt{2}}{\pi k_{\rm z}R \tilde{\eta}_{\rm c}},
\end{equation}
with $\tilde{\eta}_{\rm c}$ the Cowling's diffusion coefficient in dimensionless form. 

As with previous damping models, we infer the parameters associated to this model, $\boldsymbol {\theta}=\{ k_{\rm z}R,\tilde{\eta}_{\rm c} \}$, for different damping ratios, $d={\tau_{\rm d}/P}$, as observables. However, plausible values of Cowling's diffusion coefficient in threads are not well known, so we first explore those plausible values in this kind of coronal structures.

Determining the Cowling's diffusion coefficient is difficult because of its dependence on multiple unknown quantities in the form \citep{Leake2005}
\begin{eqnarray}
\tilde{\eta}_{\rm c}=\frac{\eta_{\rm c}}{v_{\rm Ap}R} \mbox{ where }
\eta_{\rm c}=\frac{B^2\chi^2_{\rm n}}{\mu_0\alpha_{\rm n}}\\ \mbox{ and } \alpha_{\rm n}=\frac{1}{2}\chi_{\rm n}(1-\chi_{\rm n})\frac{\rho^2_{\rm p} \Sigma_{\rm in}}{m_{\rm n}}\sqrt{\frac{16k_{\rm B}T_{\rm p}}{\pi m_{\rm i}}},
\end{eqnarray}
with $\chi_{\rm n}=2-1/\tilde{\mu}_{\rm p}$,  $\tilde{\mu}_{\rm p}$ the ionisation degree of plasma, $\Sigma_{\rm in}=5\times 10^{-19}$ m$^{-2}$ the ion-neutral cross-section, $m_{\rm n}\mbox{ and }m_{\rm i}$ the neutral and ion masses, $B$ the magnetic field strength, $\rho_{\rm p}$, $T_{\rm p}$, and $R$ the density, temperature, and radius of the thread respectively.
Assuming the prominence plasma is only constituted by neutral and ionised hydrogen, $m_{\rm n}\approx m_{\rm i}$, the Cowling's diffusion coefficient can be computed as
\begin{equation}
\tilde{\eta}_{\rm c}=\frac{2.88\times 10^{-8}B\chi_{\rm n}}{R\ \sqrt{T_{\rm p}\rho_{\rm p}^3}(1-\chi_{\rm n})}.
\end{equation}
To estimate plausible limits of this parameter, we assume threads with $R=100$ km, $\rho_{\rm p}=5\times10^{-11}$ kg m$^{-3}$, $T_{\rm p}=8000$ K, $B=5$ G, and $\chi_{\rm n}$ varying according to the ionisation degree in the range $\tilde{\mu}_{\rm p}\in(0.5,1)$. Hence, the diffusion coefficient takes values in the range $\tilde{\eta}_{\rm c}\in[10^{-4},0.5]$. Considering in addition typical wave-numbers, $k_{\rm z}R\in[10^{-3},0.1]$, and plasma-$\beta\in[0.01,1]$, theoretically predicted damping ratios for the Cowling's diffusion mechanism take values in the range $\tau_{\rm d}/P \thicksim (30 - 10^7)$, which points to this mechanism being unable to provide the observed damping time scales.

\begin{figure}[!h]
	\centering
	\includegraphics[scale=0.4]{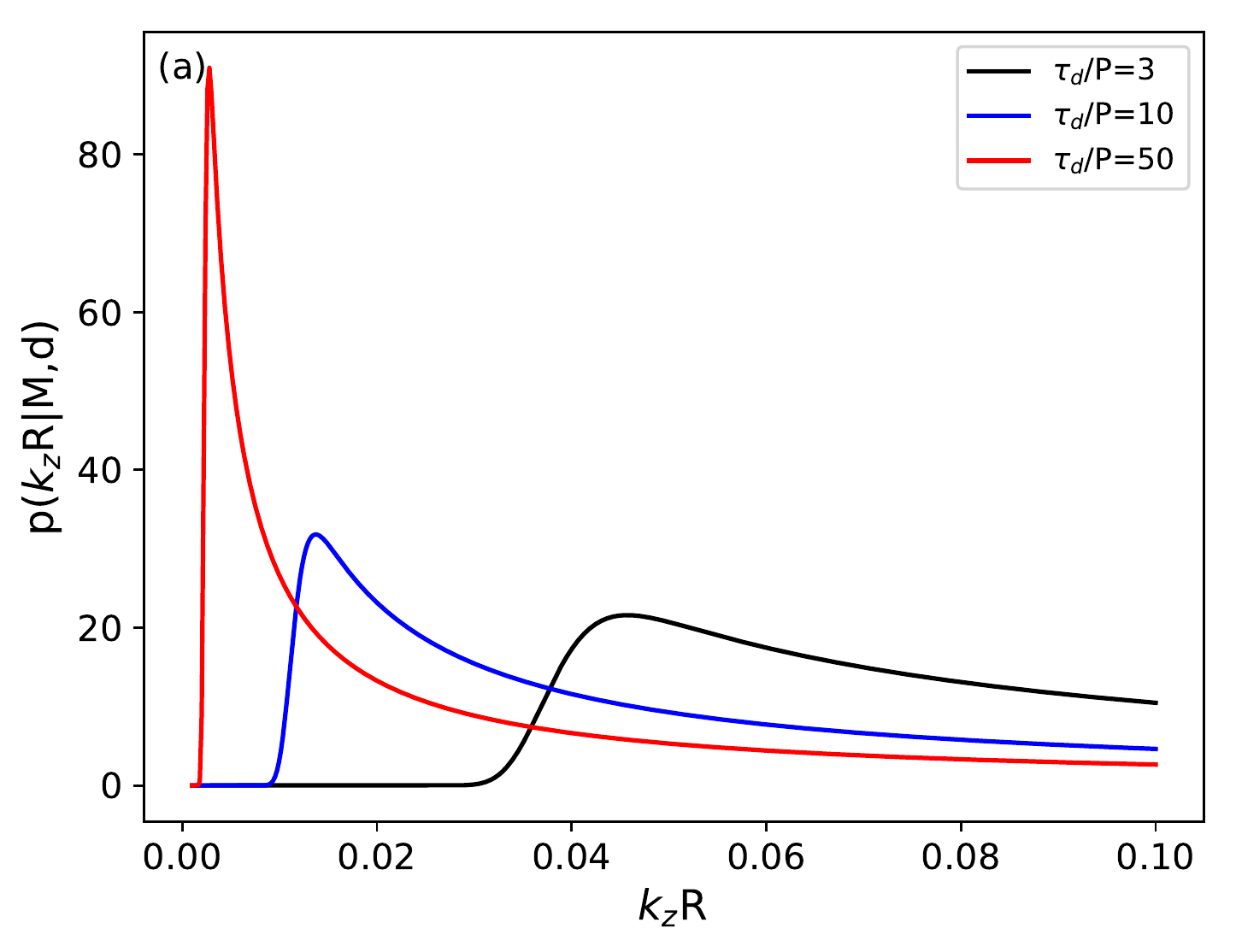}
	\includegraphics[scale=0.4]{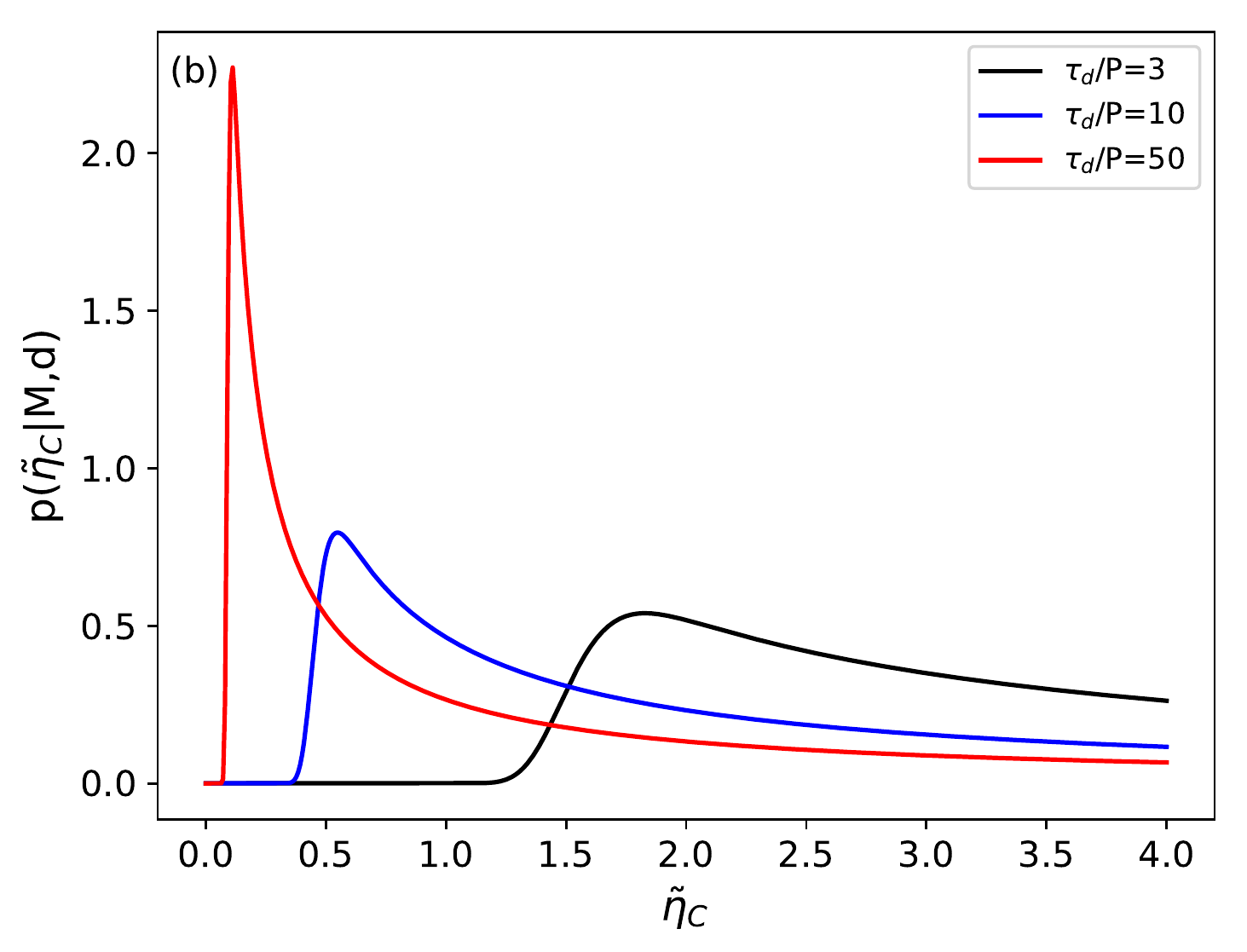}	
	\caption{Marginal posterior distributions for wave-number ($k_{\rm z}R$) and Cowling's coefficient ($\tilde{\eta}_{\rm c}$) for Cowling's diffusion. Three different damping ratios ($\tau_{\rm d}/P$) with associated uncertainty of 10\% are considered. \label{f5}}
\end{figure}

In spite of this, we performed the inference of the two relevant parameters, $k_{\rm z}R$ and $\tilde{\eta}_{\rm c}$, for three different damping ratio values. The resulting posteriors are shown in Fig.~\ref{f5}. Uniform priors have been considered. Posterior distributions can be inferred with posteriors corresponding to wave-numbers peaking at larger values for smaller damping ratios. For the inference of the Cowling's coefficient, we had to extend the range of considered values up to $\tilde{\eta}_{\rm c}=4$ to obtain posterior distributions corresponding to small damping ratios. Typical values of the diffusion coefficient seem to be adequate for damping ratios of about tens.

\subsubsection{Lengths and densities of prominence threads}\label{sec:periods}

One of the main difficulties with prominence thread seismology in comparison to coronal loop seismology is that threads show up in observations as cool and overdense plasma condensations occupying only part of a longer magnetic flux tube whose total length cannot be directly measured. The ratio of the length of the thread to the length of the tube, $L_{\rm p}/L$, is a relevant parameter upon which periods and damping times of transverse thread oscillations depend \citep{Soler2010,Arregui2011a}.

It was first suggested by \citet{Diaz2010} that the ratio between the period of the fundamental transverse kink mode to twice that of the first overtone period can be used as a tool for prominence seismology. Their analysis was based on the configuration by \citet{Diaz2002} in which the thread consists of cool material with density $\rho_{\rm p}$ occupying a length $L_{\rm p}$ embedded in a longer flux tube of length L with coronal density $\rho_{\rm c}$. In the low frequency limit ($\omega L/v_{\rm Ap} << \pi/2$), the period ratio can be cast as a function of the density contrast, $\rho_{\rm p}/\rho_{\rm c}$, and the length of the thread in units of the total length, $L_{\rm p}/L$, as
\begin{equation}
\label{eq14}
\frac{P_1}{2P_2}\approx \frac{1}{2\sqrt{L_{\rm p}/L}}\sqrt{\frac{1+L_{\rm p}/L (3f^2-1)}{1+L_p/L (f^2-1)}},
\end{equation}
with $f=\sqrt{(\rho_{\rm p}/\rho_{\rm c} +1)/2}$. If we further assume that $\rho_{\rm p}/\rho_{\rm c}>>1$, the period ratio can be approximated as 
\begin{equation}
\label{eq15}
\frac{P_1}{2P_2}\approx \sqrt{\frac{3}{4L_{\rm p}/L}}.
\end{equation}
As shown by \citet{Diaz2010}, the fundamental mode satisfies the condition of low frequency in Eq. (\ref{eq15}) but the first overtone does not. For this reason, \citet{Diaz2010} considered the next term in their  frequency  series expansion, obtaining the following expression for the period ratio
\begin{equation}
\label{eq16}
\frac{P_1}{2P_2}\approx \sqrt{\frac{3}{4L_{\rm p}/L}}\sqrt{\frac{1+\sqrt{(1+L_{\rm p}/3L)/(1-L_{\rm p}/L)}}{1+\sqrt{(9/5-L_{\rm p}/L)/(1-L_{\rm p}/L)}}}.
\end{equation}

For our Bayesian inversion we consider the period ratio, $d=P_1/2P_2$ as observable and Eqs.~(\ref{eq15}) and (\ref{eq16}) as theoretical predictions. Then, the inference of the proportion of the total length occupied by the thread, $L_{\rm p}/L$, can be attempted. A uniform prior is applied to this analysis for $L_{\rm p}/L\in(0,1)$ which includes values in between the two limiting cases of $L_{\rm p}/L=0$ (the thread does not exist) and $L_{\rm p}/L=1$ (the thread occupies the full length of the tube). For this range, Eq.~(\ref{eq16}) predicts period ratios in between 0.9 and $\infty$.  

\begin{figure}[!h]
	\centering
	\includegraphics[scale=0.5]{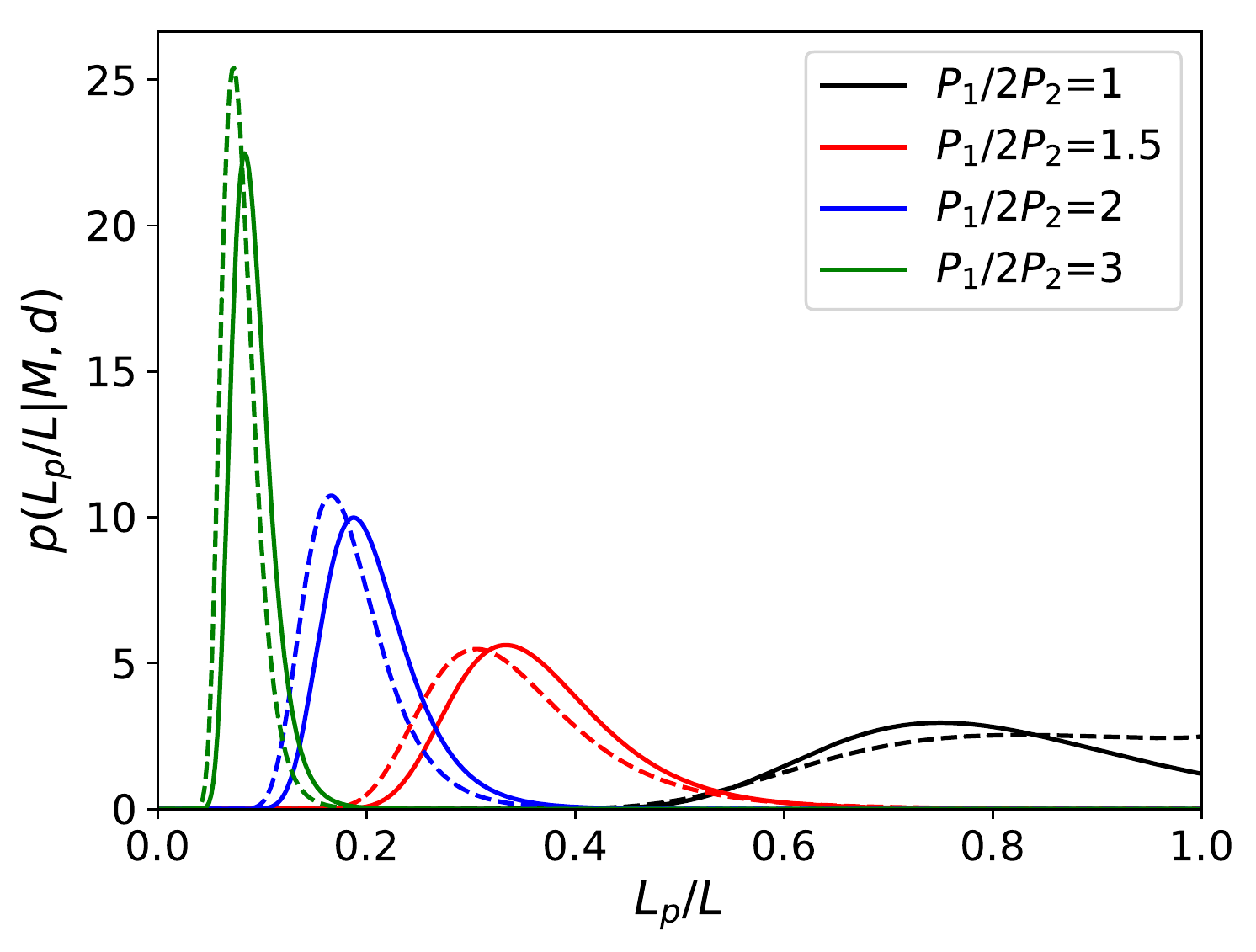}	
	\caption{Posterior distributions of proportion of thread ($L_{\rm p}/L$) considering Eq.~(\ref{eq15}) (continuous line) and Eq.~(\ref{eq16}) (dashed line), in the short frequency limit. Observable values of $P_1/2P_2$ with an associated uncertainty of 10\% have been considered. \label{f6}}	
\end{figure}

Figure~\ref{f6} shows the inferred marginal posterior distributions of $L_{\rm p}/L$ for four values of the observable period ratio and using both approximations given by Eqs.~(\ref{eq15}) and (\ref{eq16}). The larger the period ratio is, the shorter are the inferred values of $L_{\rm p}/L$. When comparing the results using both approximations, we find that  posterior distributions obtained from the Eq.~\ref{eq16} peak at slightly smaller values of the period ratio than those acquired by using the most simple equation. This means that including one more term in the series expansion of the direct problem affects the inferred value of the length of the thread. Regarding real observations, \citet{Lin2007} reported a possible detection of multiple harmonic oscillations, with a period ratio $P_1/2P_2\sim2.22$. Although the reliability of this observation is questionable, seismology applications based on this event have been presented by \citet{Soler2015} and \citet{Arregui2015b}. If we believe in the reliability of the observational estimate, this leads to an inferred value of $L_{\rm p}/L=0.16 \pm 0.03$, when an uncertainty of 10\% in the period ratio is considered.

In addition to the limitation of being only valid in the low frequency limit, Eqs.~(\ref{eq15}) and (\ref{eq16}) are not very good approximations to the period ratio for short threads (see Figure 3 of \citealt{Diaz2010}). For the short thread limit, \citet{Diaz2010} derived an expression for the period ratio of the form
\begin{equation}
\label{eq17}
\frac{P_1}{2P_2}\approx 1+(f^2-2)\frac{L}{L_{\rm p}}-(f^2+1)\left(\frac{L}{L_{\rm p}}\right)^2.
\end{equation}

\begin{figure*}[!h]
	\centering
	\includegraphics[scale=0.44]{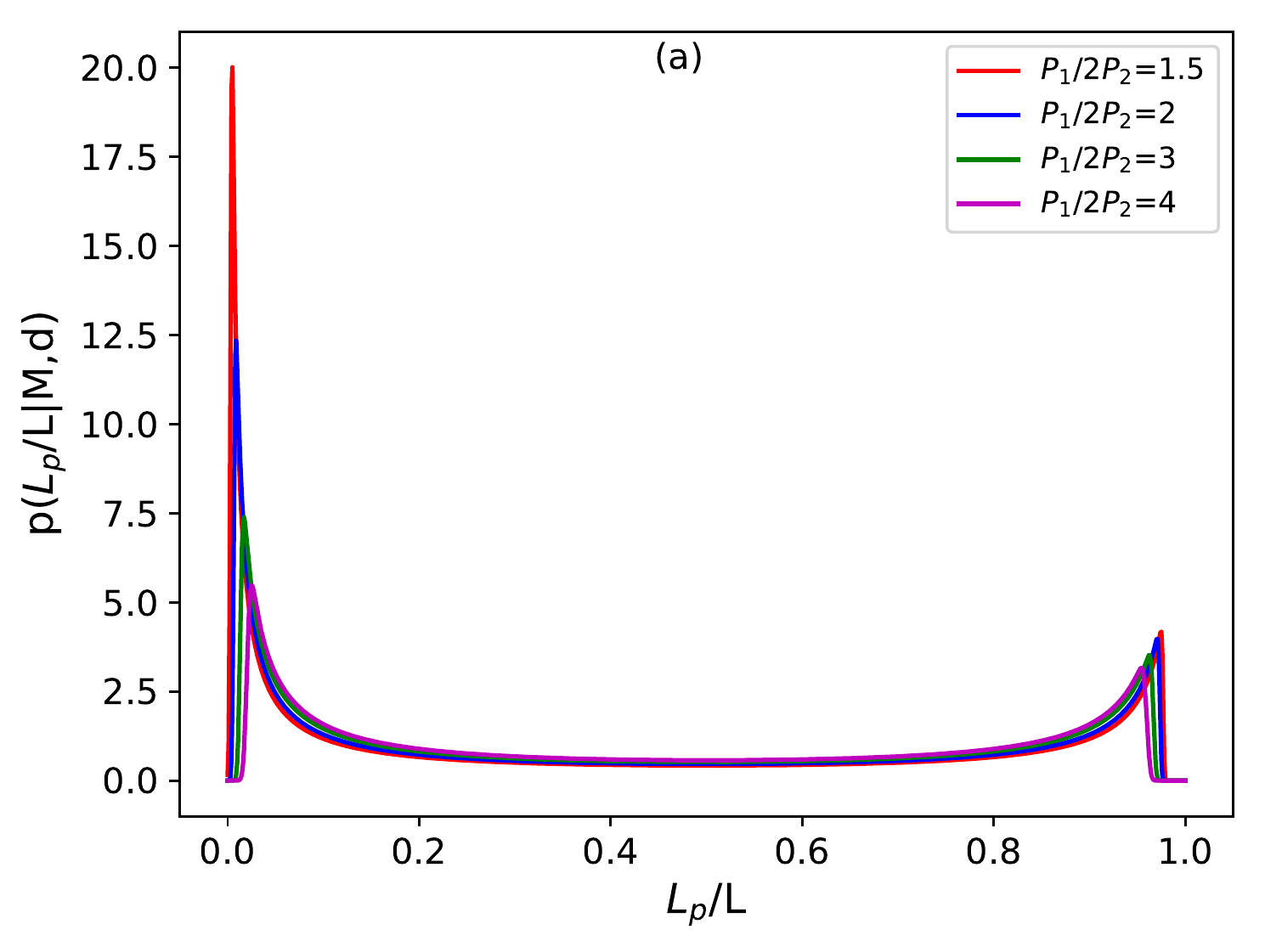}	
	\includegraphics[scale=0.44]{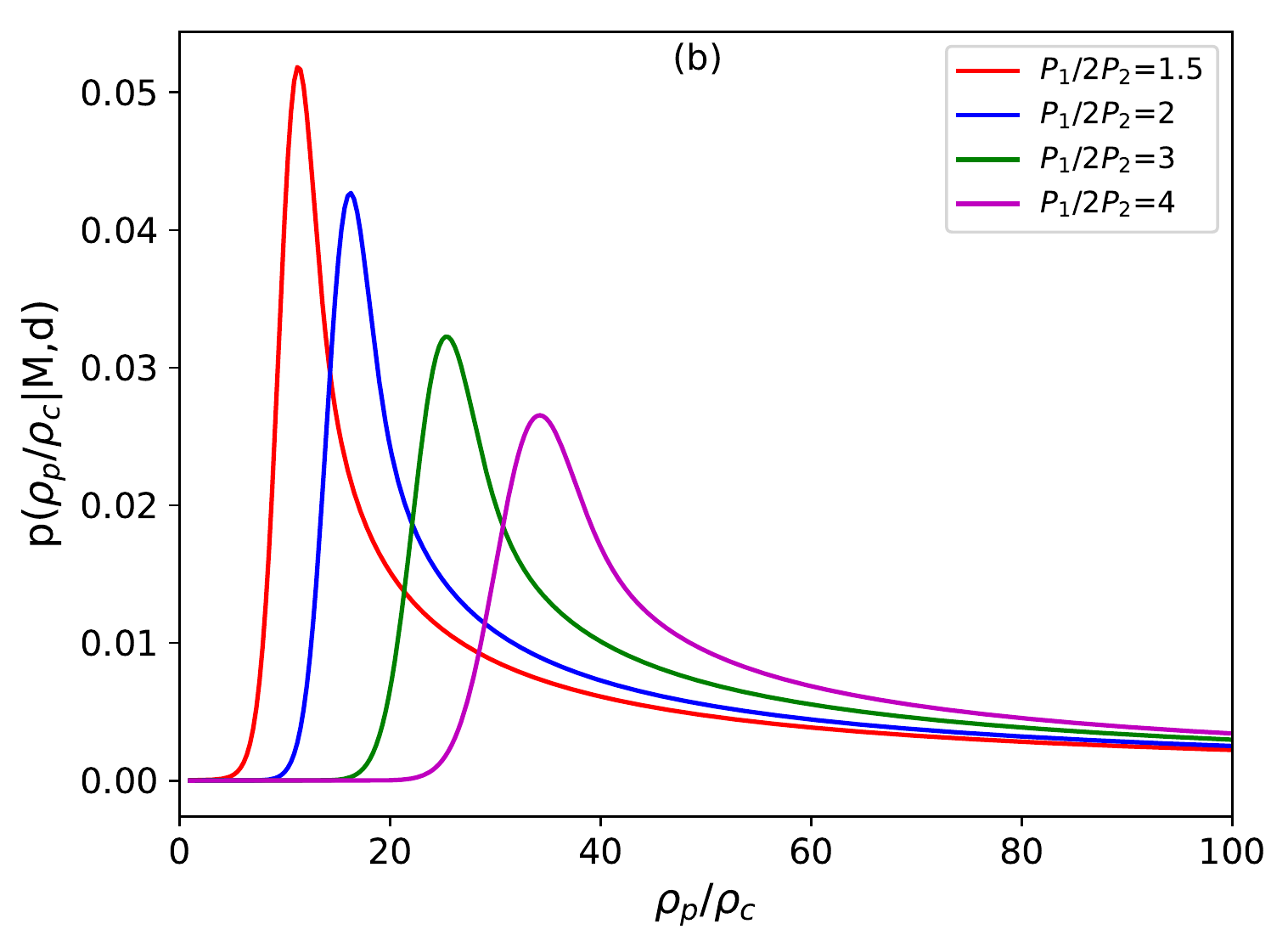}\\
	\includegraphics[scale=0.44]{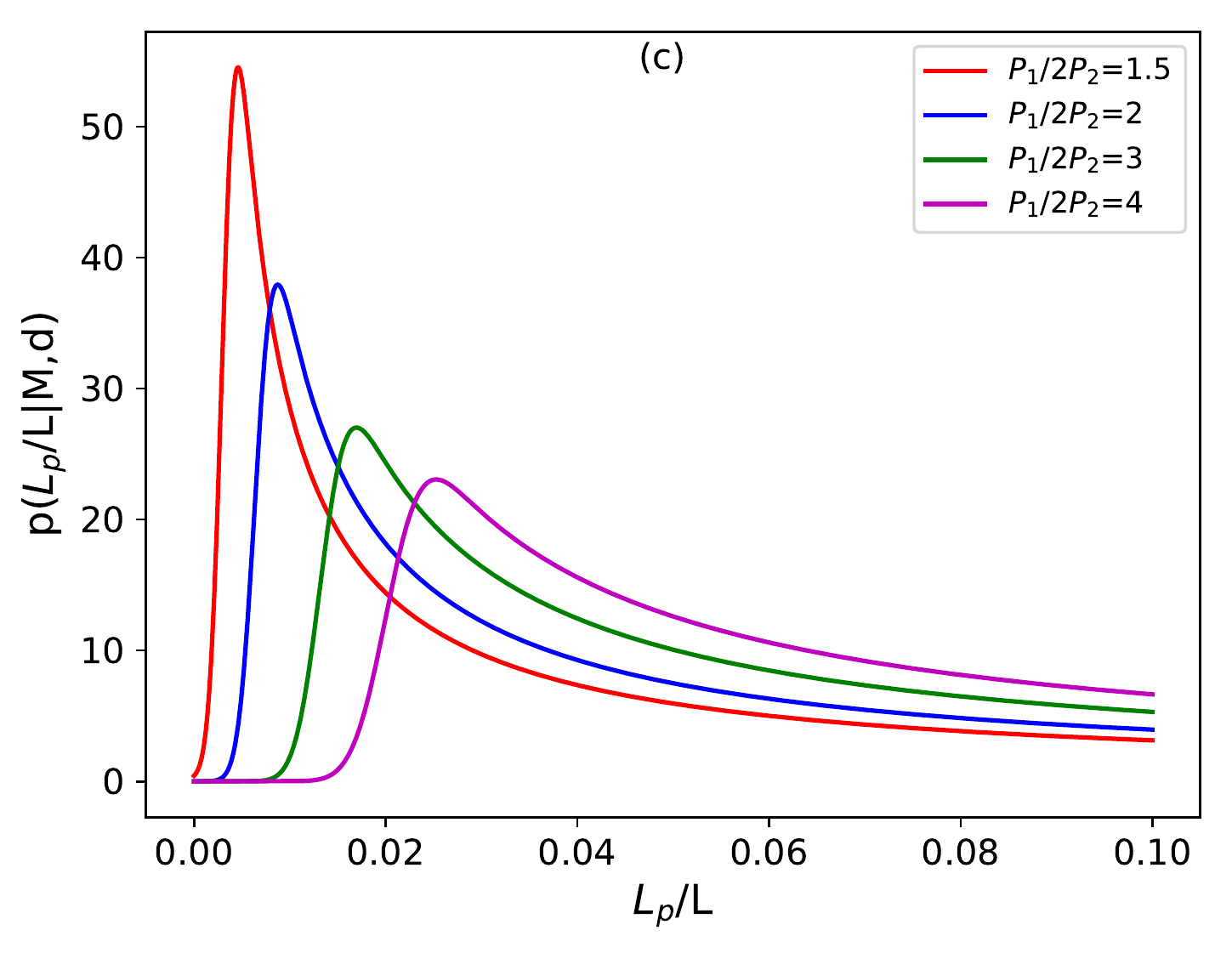}	
	\includegraphics[scale=0.44]{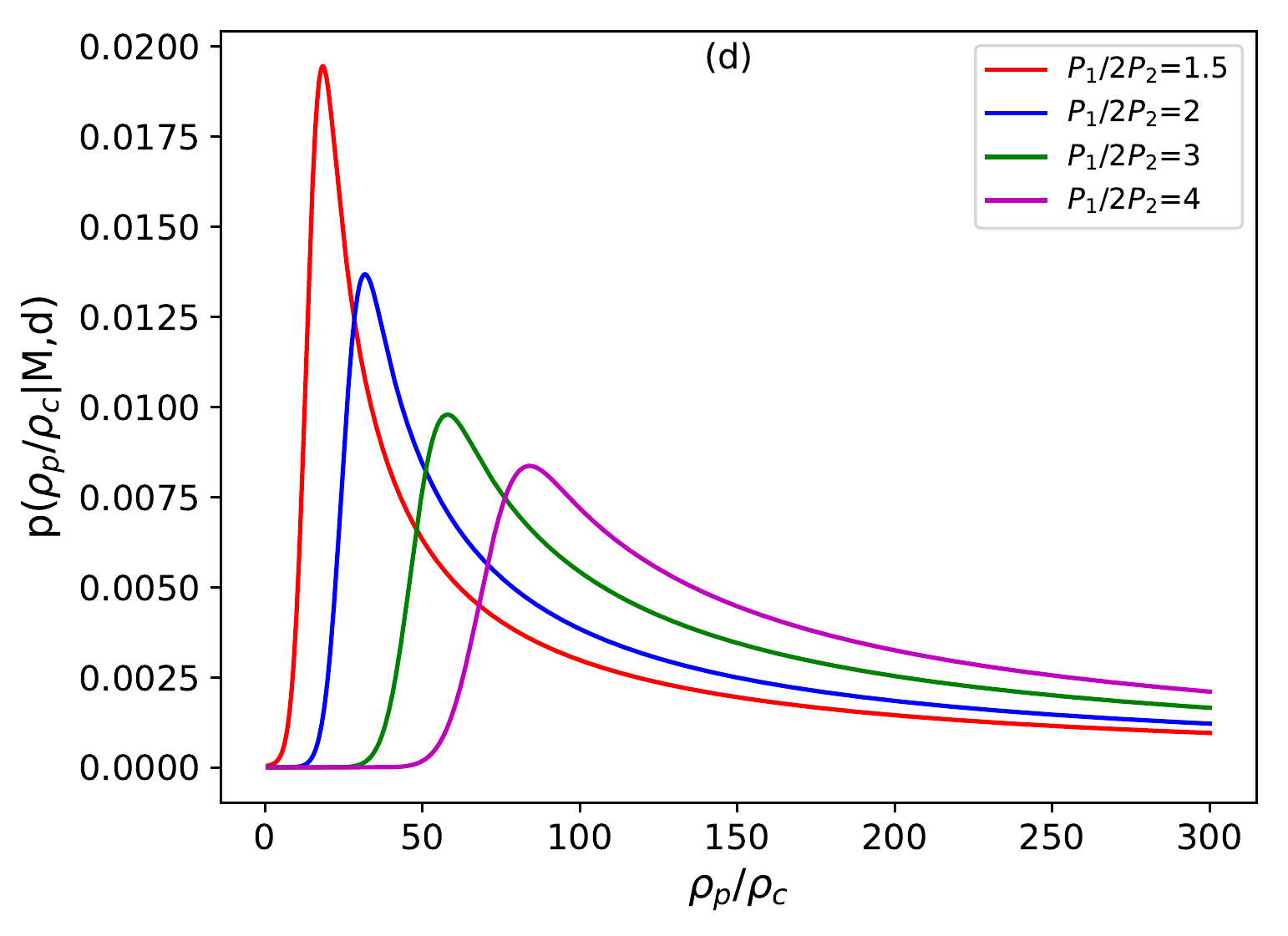}
	\caption{Posterior distributions of the thread length ($L_{\rm p}/L$) and the density contrast ($\rho_{\rm p}/\rho_{\rm c}$) for several values of the period ratio ($P_1/2P_2$) with an associated uncertainty of 10\%. Top panels correspond to use Eq.~(\ref{eq17}) with $L_{\rm p}/L\in[0,1]$ and $\rho_{\rm p}/\rho_{\rm c}\in[1.1,100]$. Bottom panels show results of considering short thread limit using the same equation than top panels with $L_{\rm p}/L\in[0,0.1]$ and $\rho_{\rm p}/\rho_{\rm c}\in[1.1,300]$.\label{f7}}	
\end{figure*}

Using the same inference procedure as before, we now use Eq.~(\ref{eq17}) to infer the two parameters $\boldsymbol {\theta}=\{ L_{\rm p}/L, \rho_{\rm p}/\rho_{\rm c}\}$ from the observable period ratio. Uniform priors are used for the length of the thread in the range of $L_{\rm p}/L\in[0,1]$ as a first approximation to sample all plausible values and for the density contrast $\rho_{\rm p}/\rho_{\rm c}\in[1.01,300]$ to account for the large density contrast typical of prominence plasmas. Figure~\ref{f7}a and b show the resulting posterior probabilities for both parameters, $L_{\rm p}/L$ and $\rho_{\rm p}/\rho_{\rm c}$. For all considered period ratios, both the density contrast and the length of the thread can be properly inferred. Figure~\ref{f7}a, corresponding to the length of thread, shows two clearly visible peaks for each distribution with small differences between them for different period ratio values. Small values of the theoretical period ratio are compatible with large lengths of thread and small density contrast but also with small lengths of thread and large density contrasts because of the negative sign of last term in Eq.~(\ref{eq17}). The distributions for the density contrast in Fig.~\ref{f7}b peak at small values of the parameter spread to larger values for larger period ratio values. In the considered range for this parameter, the secondary peak is not visible.

Note that Eq.~(\ref{eq17}) is only strictly valid in the short thread limit. If we constrain the possible values of $L_{\rm p}/L$ to the shorter range $L_{\rm p}/L\in[0,0.1]$, we obtain the results shown in Figs.~\ref{f7}c and d. In contrast to the tendency found for the marginal posteriors for $L_{\rm p}/L$ in the long thread approximation, the posteriors in Fig.~\ref{f7}c move towards larger values of $L_{\rm p}/L$ for larger period ratio values. A similar behaviour is obtained for the posterior distributions of density contrast which peak at larger values in comparison to those obtained considering the full range in $L_{\rm p}/L$ in the previous case. Also, a very small probability is obtained for density contrasts above 200, unless we consider wider ranges of $\rho_{\rm p}/\rho_{\rm c}$ with values from tens to thousands.

\begin{figure}[h]
	\centering
	\includegraphics[scale=0.5]{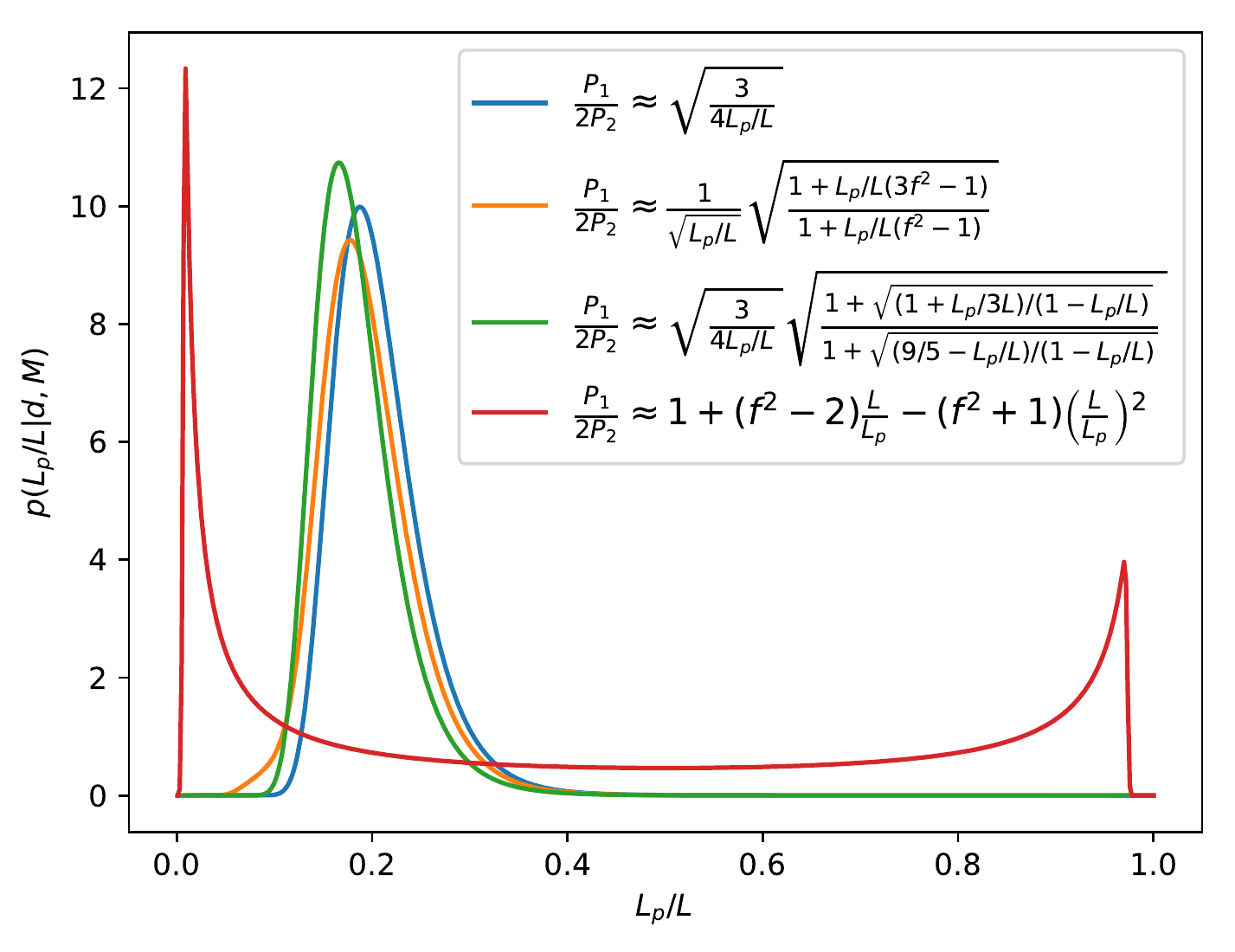}
	\caption{Comparison of posterior distributions of the thread length proportion ($L_{\rm p}/L$) obtained for  Eqs.~(\ref{eq14}), (\ref{eq15}), (\ref{eq16}), and (\ref{eq17}). One observable period ratio of $P_1/2P_2=2$ has been considered with an associated uncertainty of 10\%.\label{f8}}
\end{figure}

To analyse possible differences between several equations in this section, i.e. Eqs.~(\ref{eq14}), (\ref{eq15}), (\ref{eq16}), and (\ref{eq17}), we compute the posterior of the length of thread proportion for all plausible range of the parameter in all cases with an observable period ratio equal to $P_1/2P_2=2.0\pm 0.2$. Figure~\ref{f8} shows posterior distributions resulting from this analysis. In the long thread approximation, distributions obtained from considering different equations do not differ. However, the result differs when we consider the equation for the short thread approximation. The posterior distribution has one peak in a value next to zero differing from the other distributions. In addition, it shows a secondary peak with smaller amplitude at the opposite side of parameter range, next to 1 due to the sign of the last term in the equation.

\subsubsection{Length of flowing and oscillating threads}
\begin{figure*}[!t]
	\centering
	\includegraphics[scale=0.35]{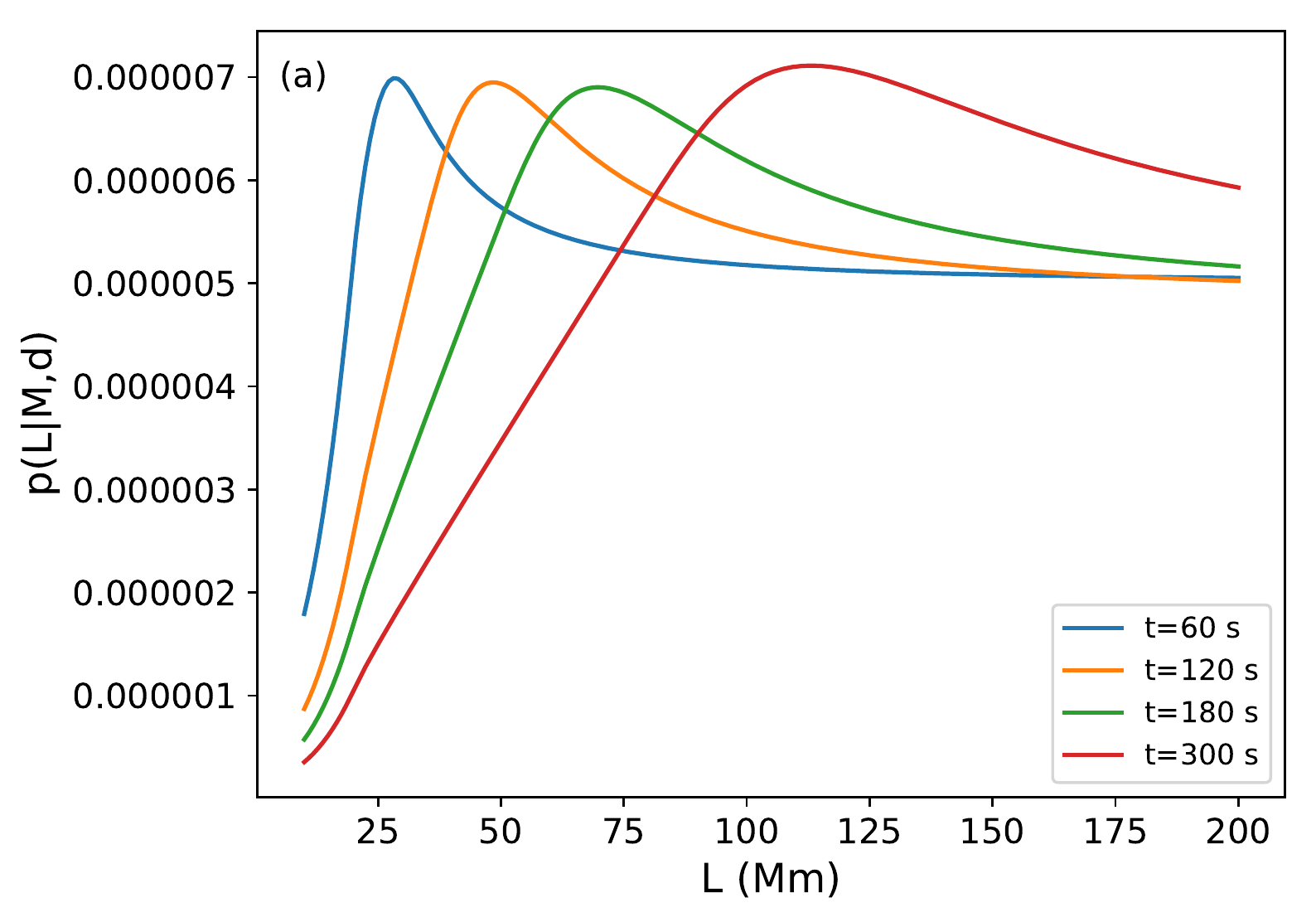}
	\includegraphics[scale=0.35]{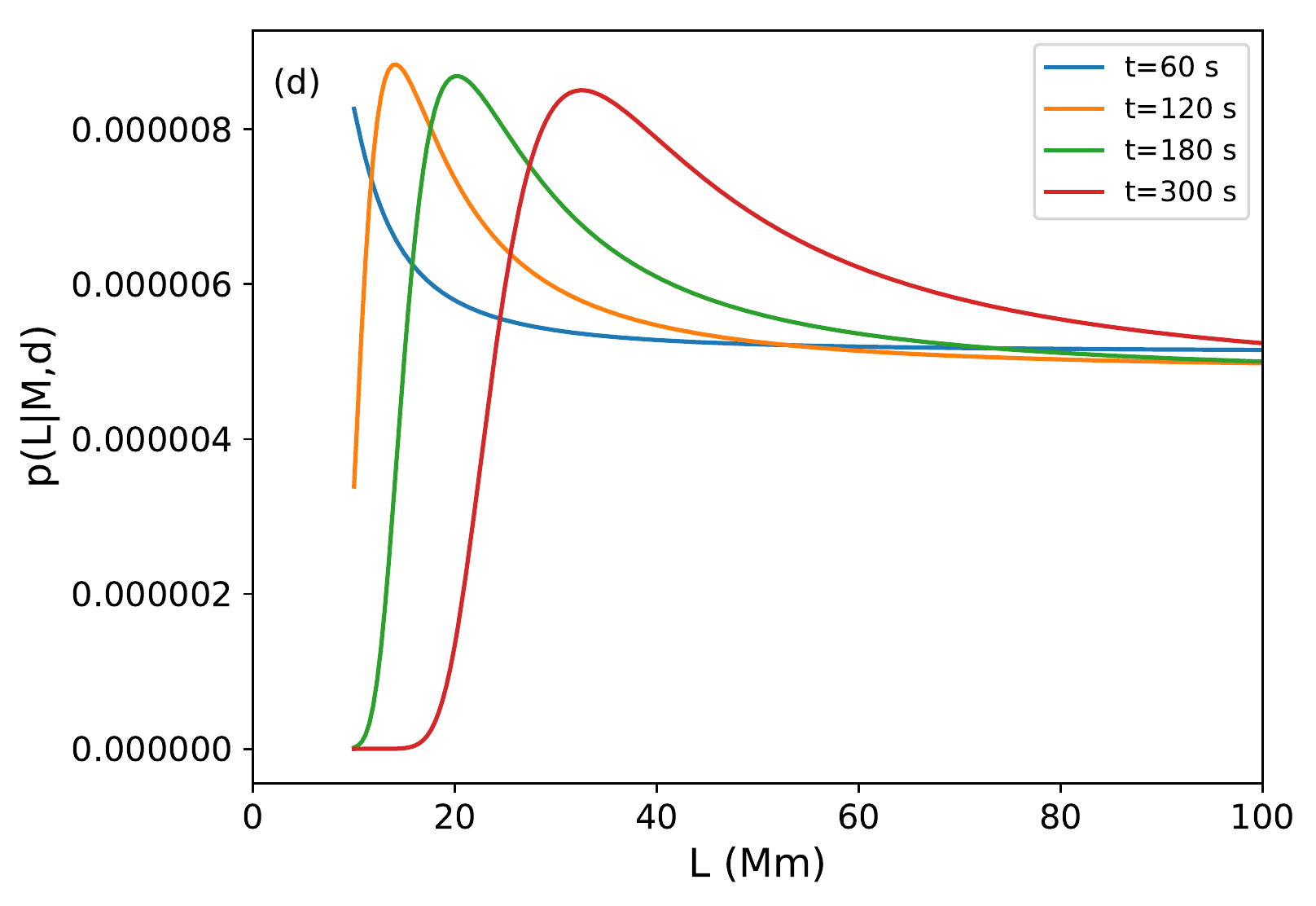}\\
	\includegraphics[scale=0.35]{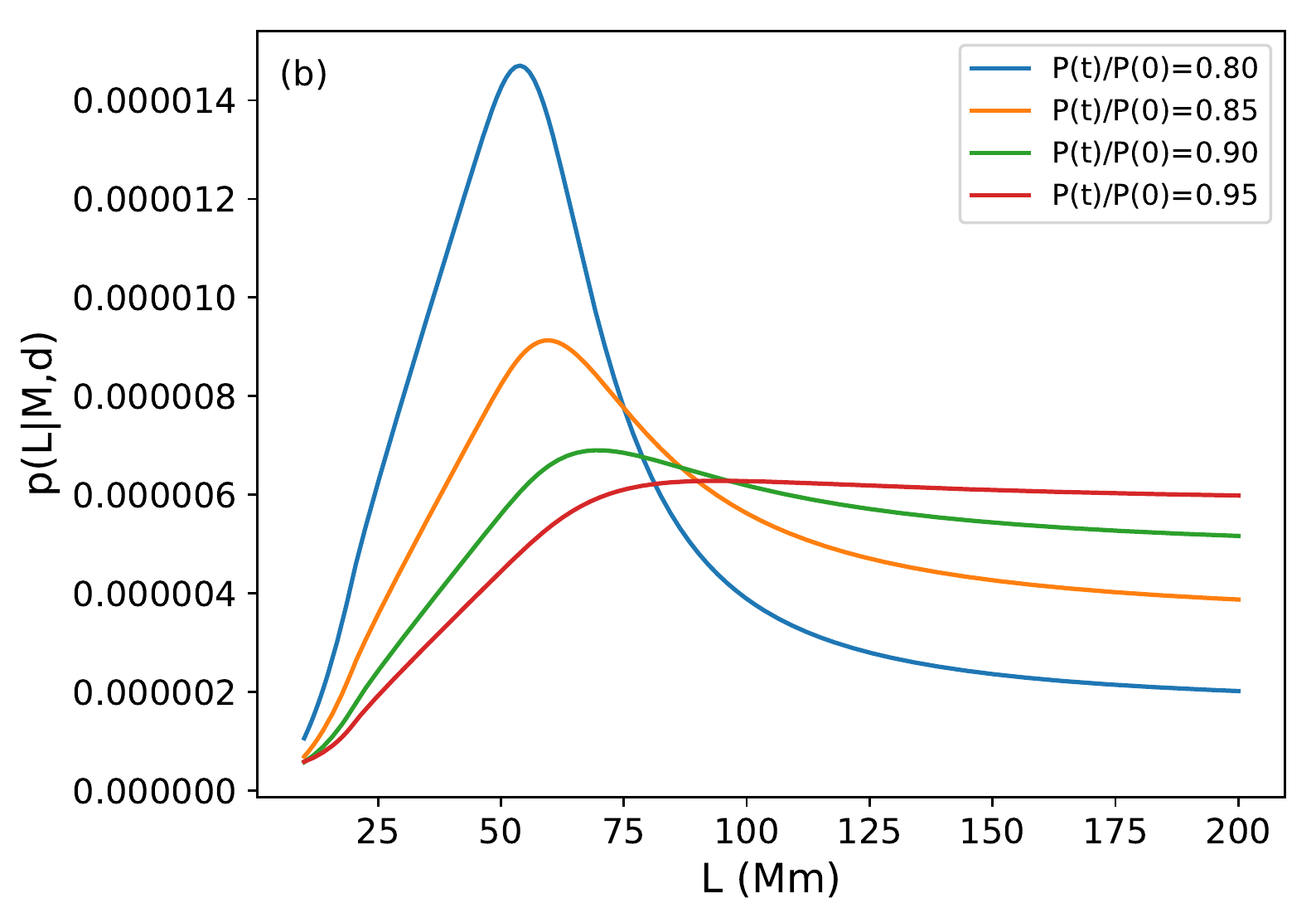}
	\includegraphics[scale=0.35]{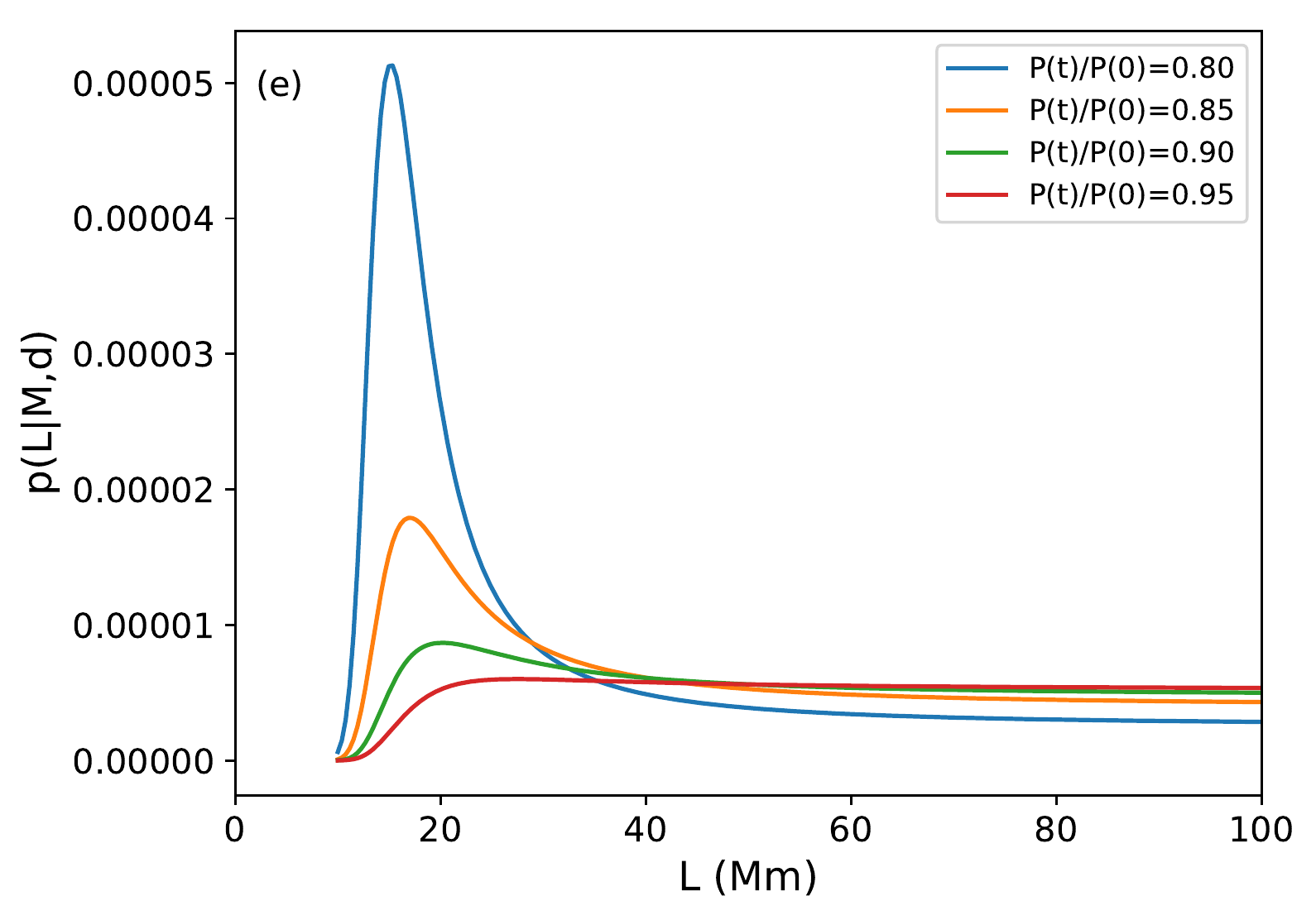}\\
	\includegraphics[scale=0.35]{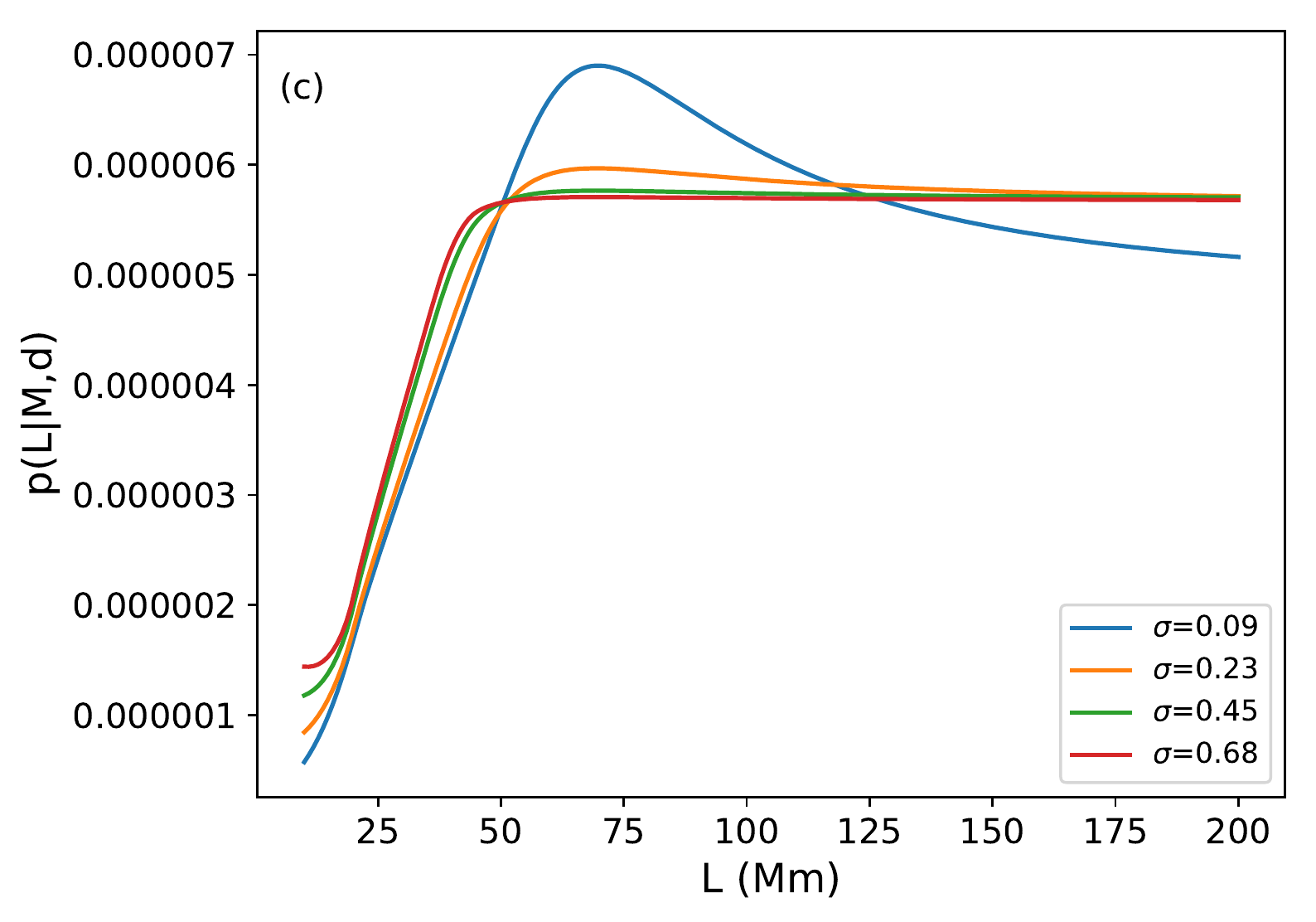}
	\includegraphics[scale=0.35]{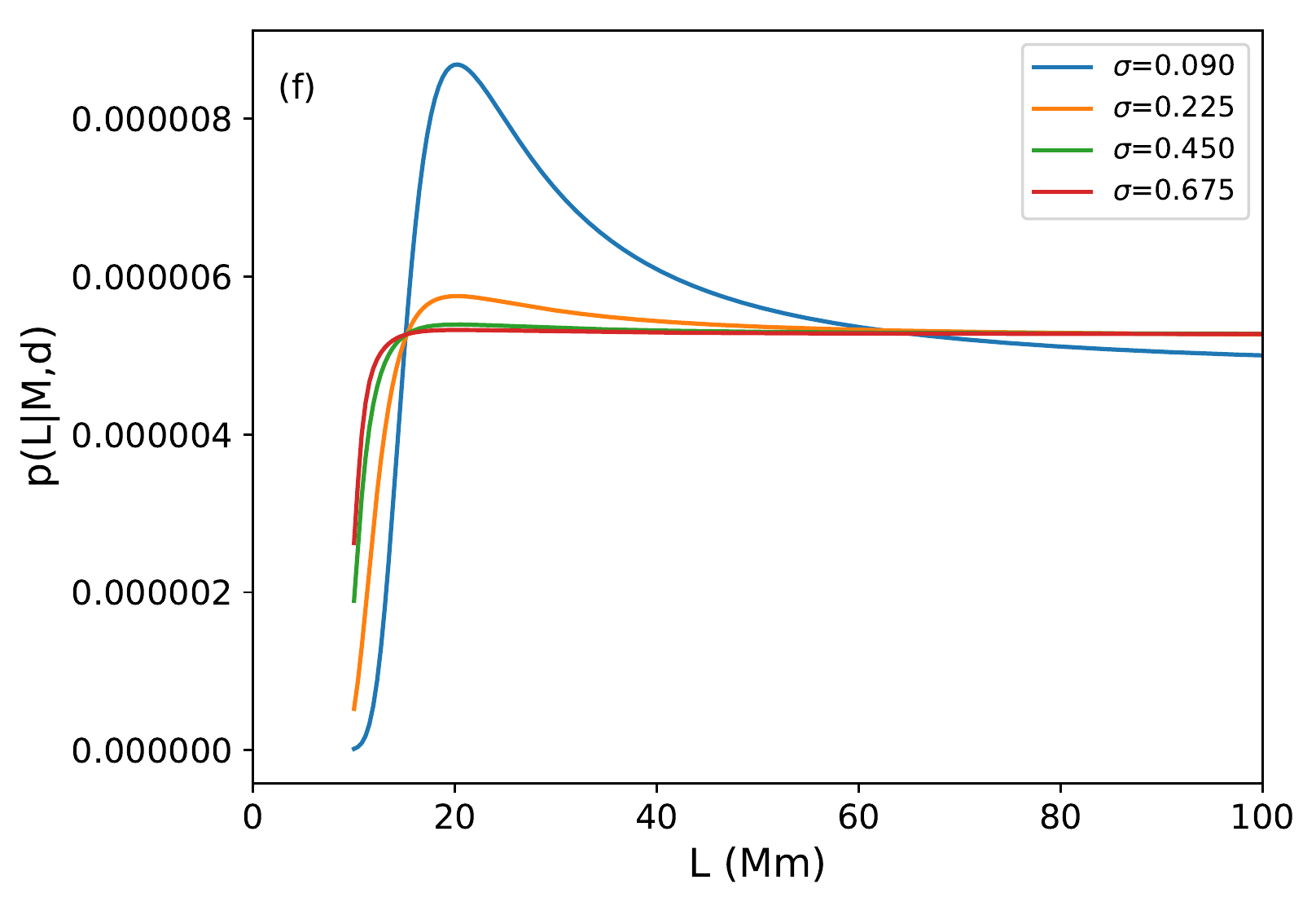}
	\caption{Posterior distributions of the total length ($L$) with uniform (left panels) and Gaussian (right panels) priors. Gaussian priors are centred in $v_0=22$ kms$^{-1}$ and $L_{\rm p}$=4 Mm with an uncertainty of 10\%. First row shows results of considering different observation times with $P(t)/P(0)=0.9$ and an associated uncertainty of 10\%. Second row shows results of considering different period ratios with $t=180$ s and 10\% as uncertainty for all observables. Third row conforms results associated to different uncertainties, $\sigma=10,25,50,75\%$, with $P(t)/P(0)=0.9$ and $t=180$ s.\label{f9}}
\end{figure*}

Prominence threads are observed flowing at the same time that they support transverse oscillations \citep{Okamoto2007,Okamoto2015}. Some theoretical models have considered the properties of transverse waves in non-static equilibria and the influence of mass flows in their oscillatory features \citep{Dymova2005}, with first applications to Hinode observations \citep{Terradas2008}.

To perform our analysis, we consider again the partially filled tube model based on \citet{Diaz2002} but the thread is permitted to flow along the magnetic field direction. In this scenario, the inclusion of a flowing thread with velocity $v_0$ produces the period of fundamental kink mode to vary in time. \citet{Soler2011} analysed the temporal evolution of the period and gave an analytic expression for the ratio between the period at any time, $t$, and the period at time $t=0$, of the form
\begin{equation}
\label{eq18}
\frac{P(t)}{P(0)}= \sqrt{1-\frac{4v_0^2t^2}{(L+\frac{1}{3}L_{\rm p})(L-L_{\rm p})}}.
\end{equation}
This expression depends on three physical quantities, the flow velocity, $v_0$, the length of the thread, $L_{\rm p}$, and the total length of the flux tube, $L$.

Considering the theoretical prediction given by Eq.~(\ref{eq18}), we first compute the posterior distribution for each model parameter $\boldsymbol{\theta}=\{L,v_0,L_{\rm p}\}$ assuming a time of observation equal to $180$ s and an observable $d=P(t=180$ s$)/P(0)=0.9$, with an associated uncertainty of 10\%. The possible values of parameters inside the square root are limited by mathematical reasons since the second term should be less than 1, so that considered ranges of parameters are $v_0\in(0,100]$ kms$^{-1}$, $L_{\rm p}\in[0,20]$ Mm, and $L\in[50,200]$ Mm, but not all combinations between them will be possible. Uniform priors have been assumed for the three quantities in a first scenario and Gaussian priors for the flux velocity and length of the thread in a second scenario. 

The resulting posterior distributions for the total length of the flux tube are presented in Fig.~\ref{f9} for the two scenarios. Different values of observation times, period ratios, and uncertainties are studied. The flow velocity and the thread length cannot be properly inferred and they are omitted for clarity. In general, distributions are not well constrained showing long tails for large values of the parameter but more defined distributions are obtained when Gaussian priors are assumed. Panels (a) and (d) show posteriors of $L$ for different times of observations. Larger values of $L$ are more plausible for larger times. A lower limit of $L$ could be inferred in the second scenario. Panels (b) and (e) present posterior distributions for different period ratio values. Smaller values of $L$ are more likely and more defined distributions are obtained with smaller period ratio values. Regarding the influence of considering different uncertainties in measurements, panels (c) and (f) show that posterior distributions for the total length of the flux tube do not change significantly except for the smallest value of the uncertainty. The length of the flux tube cannot be well inferred, unless the uncertainty associated to the period ratio measurement is very small.

After analysing the general case, we focus on applying the same model to particular real observations reported by \citet{Okamoto2007}. These authors report on the simultaneous presence of flowing and transverselly oscillating threads observed with Hinode. Although period variations in time were not reported, we assume hypothetical period changes to exemplify a possible application to infer the total length of the flux tube from this kind of observations. Table~\ref{tab4} gives a summary of the thread lengths and flow velocities reported by \cite{Okamoto2007}. In our inference, the observable and its uncertainty remain the same as in the previous scenario but Gaussian priors are used for the length of the thread and the flow velocity, with Gaussians centred at those values measured by \citet{Okamoto2007}. Uncertainties are not given in their work, so that we have considered uncertainties of 10\% for all measurements. Regarding the total length of the flux tube, a uniform prior is assumed.
\begin{figure}[!t]
	\centering
	\includegraphics[scale=0.5]{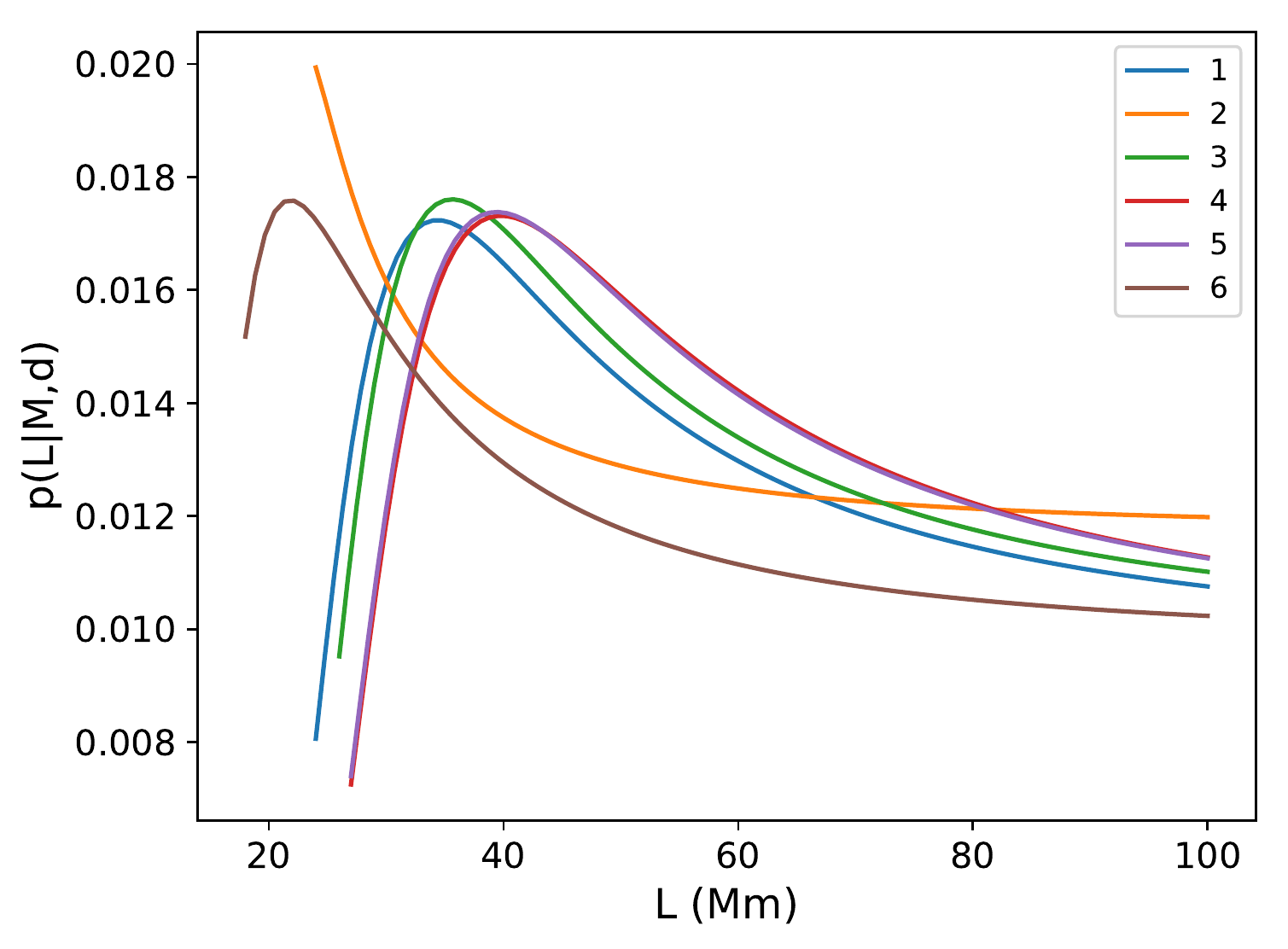}
	\caption{Posterior distributions of the total length of threads analysed by \citet{Okamoto2007}. An observation time of $t=180$ s and a ratio of 0.9 have been assumed with an associated uncertainty of 10\%. \label{f10}}	
\end{figure}

Results for the inferred distributions and summary values of median with errors for the total length for all six threads are shown in Fig.~\ref{f10} and Table~\ref{tab4}, third column. The total length of the flux tube cannot be properly inferred because the posterior distributions show long and high tails at the right-hand side of the distributions. However, all six distributions show a common tendency to peak at values of the length in between 20 Mm and 40 Mm approximately, except for thread numbers 2 and 6 whose posteriors cannot be inferred because of limitations imposed by Eq.~(\ref{eq18}). Analysing the median values in Table~\ref{tab4}, the total length takes values roughly in between 20 and 90 Mm within the errors, being this last value next to values reported in some previous studies which fixed a minimum value of 100 Mm \citep{Soler2010,Terradas2008}. Hence, $L_{\rm p}$ represents less than 10\% of the total flux tube.
\begin{table}[h]
	\caption{Threads data analysed by \citet{Okamoto2007}. Columns contain the number of thread (\#), the observed thread length ($L_{\rm p}$), the measured velocity of flow ($v_0$), and the median values of posterior distributions of the total length of the flux tube ($L$) with a credible interval of 68\%.\label{tab4}} 
	
	\centering 
	\begin{tabular}{c c c c} 
		\hline\hline 
		Thread & $L_{\rm p}$ & $v_0$ & L \\ 
		\# & (Mm) & (kms$^{-1}$) & (Mm) \\
		\hline 
		1 & 3.6& 39& $58^{+28}_{-22}$\\\\
		2 & 16& 15& $59^{+28}_{-25}$\\\\
		3 & 6.7& 39& $59^{+28}_{-22}$\\\\
		4 & 2.2& 46& $60^{+27}_{-21}$\\\\
		5 & 3.5& 45& $60^{+27}_{-21}$\\\\
		6 & 1.7& 25& $54^{+31}_{-26}$\\
		\hline 
	\end{tabular}
\end{table}

It is worth to mention that the flow speeds by \citet{Okamoto2007} might be apparent and due to the presence of intensity variations from heating/cooling processes. Also, the authors did not consider any uncertainty coming from e.g., line of sight projection effects.

\subsection{Model comparison}
The use of Bayesian techniques provides information about model parameters and further permits to compare the capability of alternative models in explaining observations. In this section, we present results from two applications of model comparison. We first compare the plausibility of the three damping models considered in Section~\ref{sec:damping}. Then, we do the same with the two alternative period ratio approximations in the short and long thread limits described in Section~\ref{sec:periods}. 

\subsubsection{Damping mechanisms} 

The causative mechanism of damping of transverse oscillations in threads remains unknown. A simple evaluation of the capacity of each damping mechanism to reproduce observed damping ratios can be obtained by inserting typical values for the model parameters in Eqs.~(\ref{eq8}), (\ref{eq9}), and (\ref{eq10}), corresponding to resonant absorption in the Alfvén continuum, resonant absorption in the slow continuum, and Cowling's diffusion, respectively. The ranges so obtained and already discussed in Section~\ref{sec:damping} point towards the Alfvén resonance as the most plausible mechanism. However, one must be aware that those so-called typical values are highly uncertain and cannot be measured directly. Therefore, different combinations of typical values would lead to different predictions of $\tau_{\rm d}/P$ for each model, which could have a large variability. It is this variability what Bayesian model comparison permits to analyse, by studying how often different model parameter combinations lead to a given observable.
\begin{figure}[!h]
	\centering
	\includegraphics[scale=0.33]{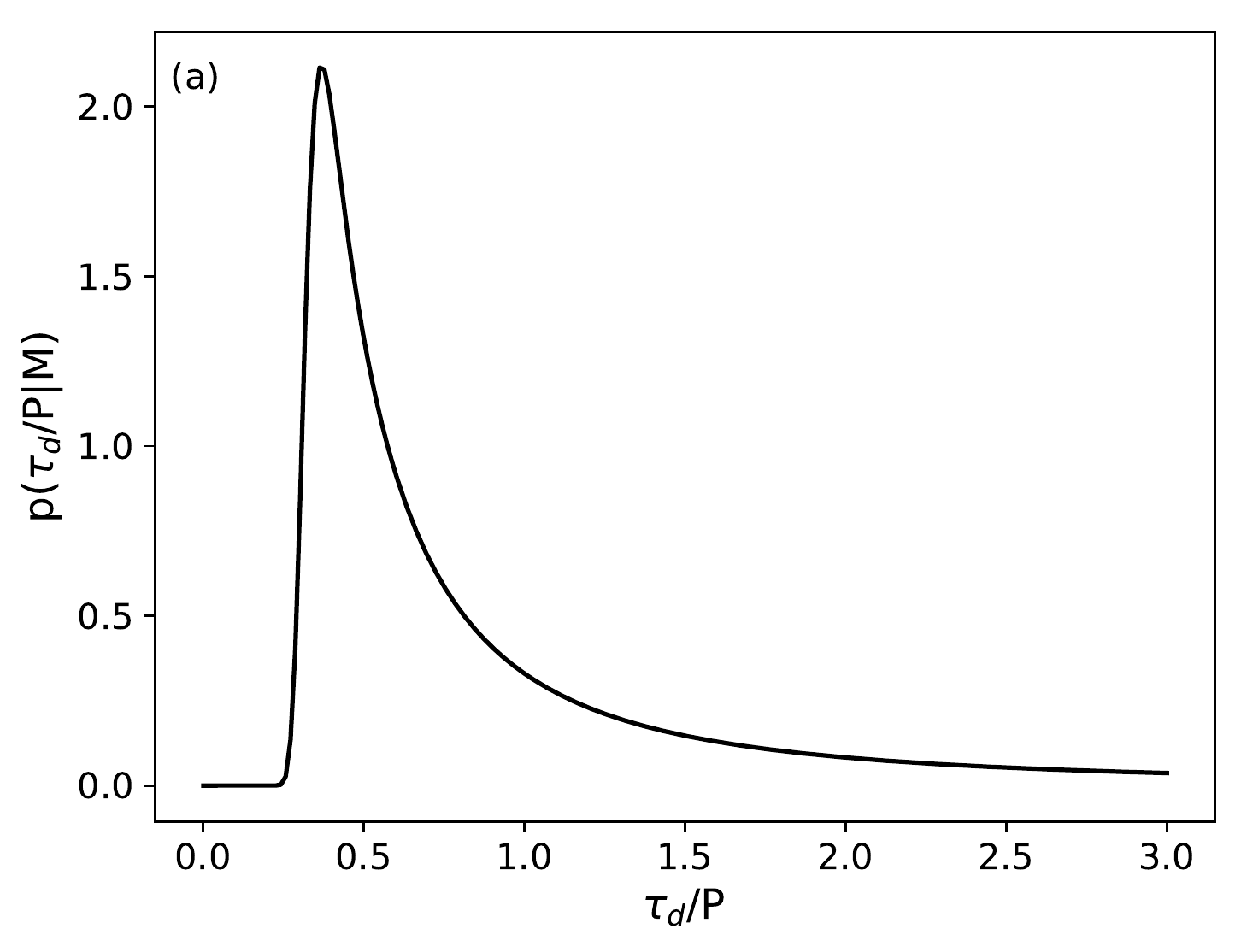} 
	\includegraphics[scale=0.33]{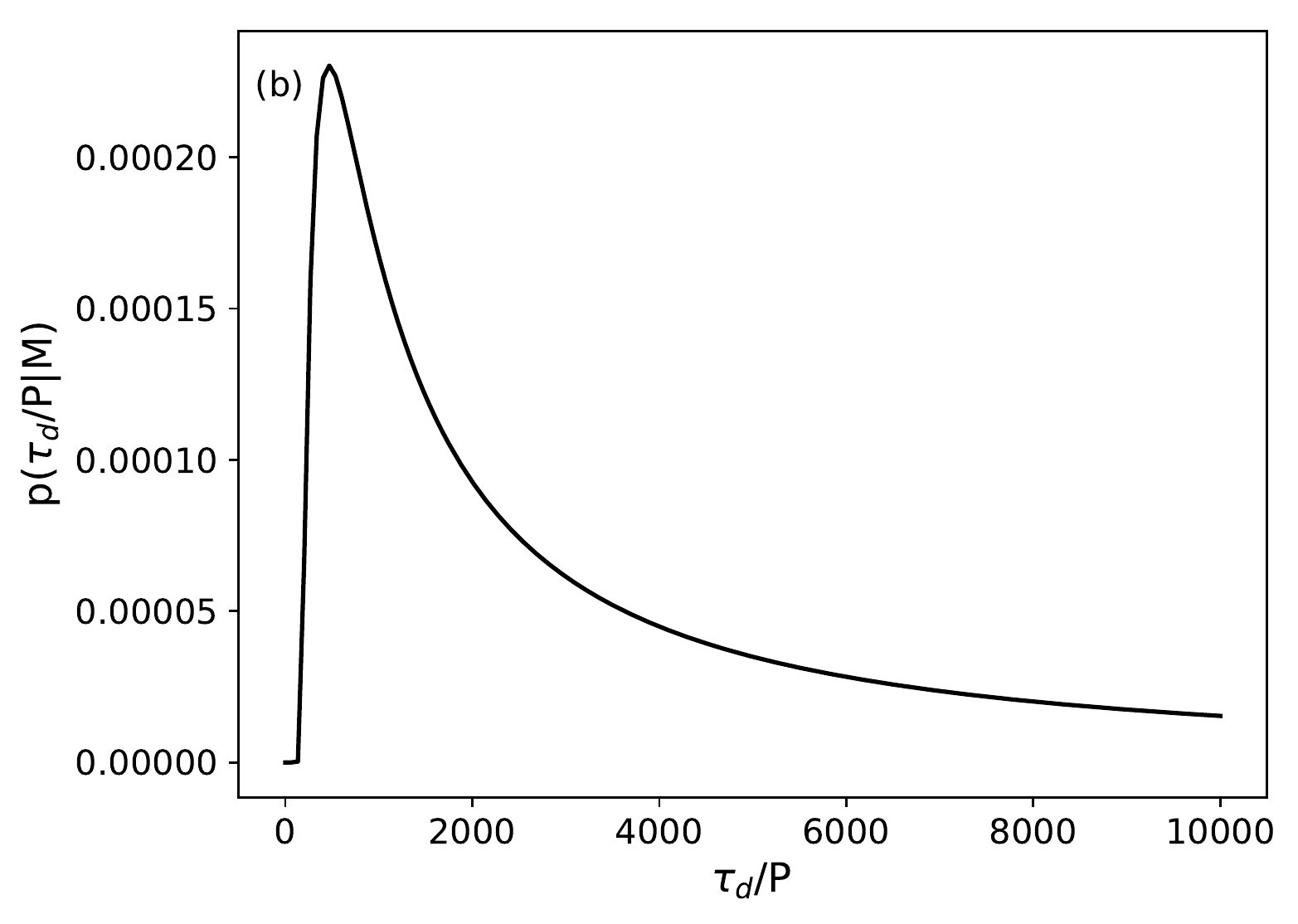}
	\includegraphics[scale=0.33]{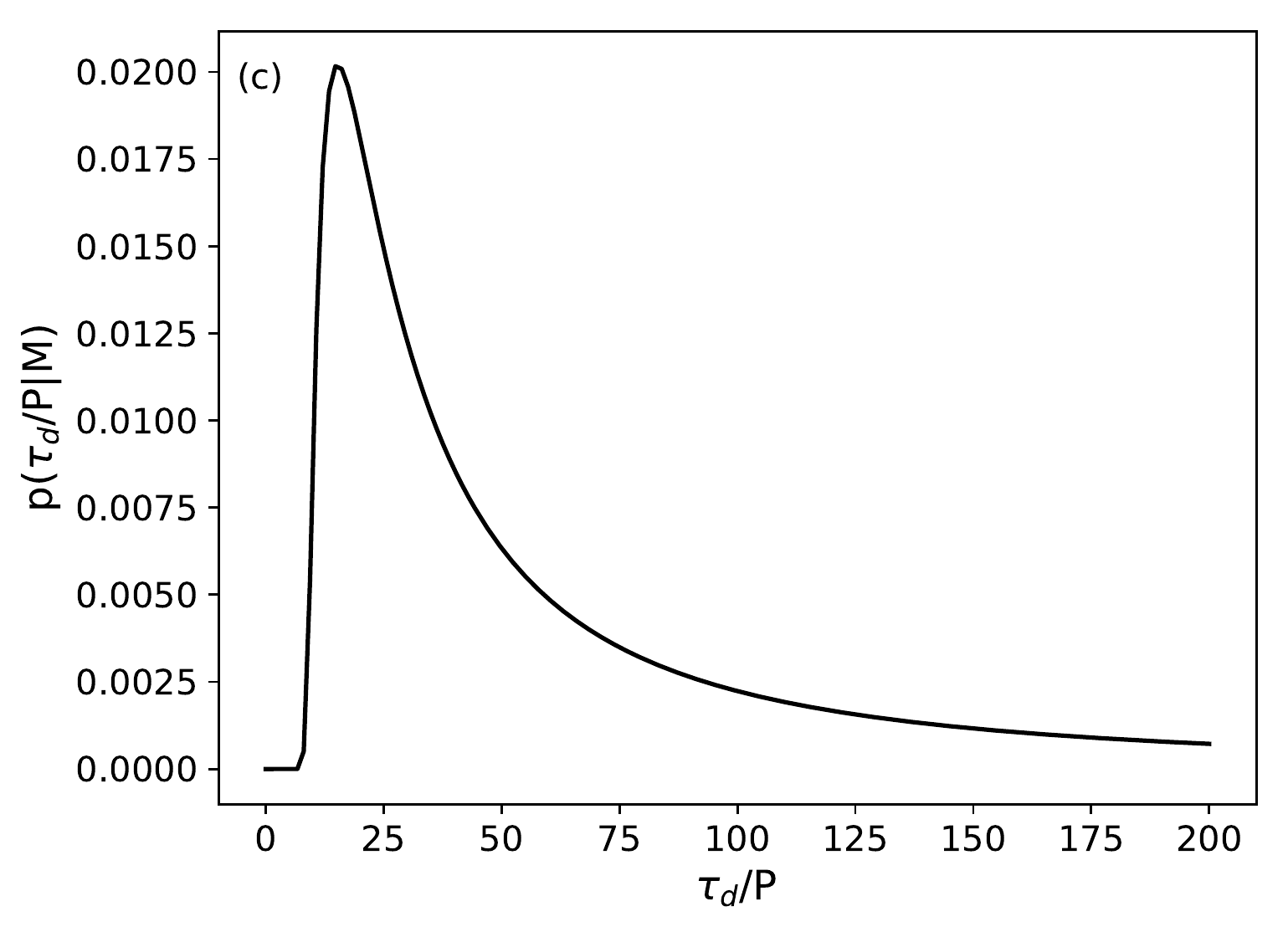}
	
	\caption{Marginal likelihoods for the three selected damping mechanisms. (a) Resonant absorption in the Alfvén continuum. (b) Resonant absorption in the slow continuum. (c) Cowling's diffusion. $\tau_{\rm d}/P$ values are indicated in the $x$-axis. An uncertainty of 10\% has been used. \label{f11}}	
\end{figure}

We first use the marginal likelihood, as defined in Eq.~(\ref{eq2}), to calculate the likelihood of a given model to reproduce a given observed damping ratio, considering how different combinations of model parameters contribute to theoretical predictions that are nearby observed data. This ''nearby'' will also depend on the error on the measured damping ratio. In our case, we considered a 10 \% uncertainty on the observable, $d=\tau_{\rm d}/P$. Uniform and independent priors for all parameters of the damping models have been contemplated. The values for these parameters are enclosed in the same plausible ranges for prominence conditions used in the inference analysis. 

Figure~\ref{f11} shows marginal likelihoods corresponding to each damping mechanism, as a function of the observable damping ratio. Resonant absorption in the Alfvén continuum (Fig.~\ref{f11}a) has the largest plausibility for very strong damping regimes, with values of $\tau_{\rm d}/P$ even below 1. The marginal likelihood for resonant absorption in the slow continuum (Fig.~\ref{f11}b) spreads over much larger damping ratio values, peaking at a damping ratio around 470. Finally, the marginal likelihood for Cowling's diffusion  (Fig.~\ref{f11}c) is more compatible with observed damping ratios in the range $5\times10^6-5\times10^7$. Hence, each mechanism seems to explain better different ranges of $\tau_{\rm d}/P$, but only the Alfvén resonant absorption shows a marginal likelihood that has its largest values for damping ratios compatible with those observed. 

\begin{figure}[!h]
	\centering
	\includegraphics[scale=0.35]{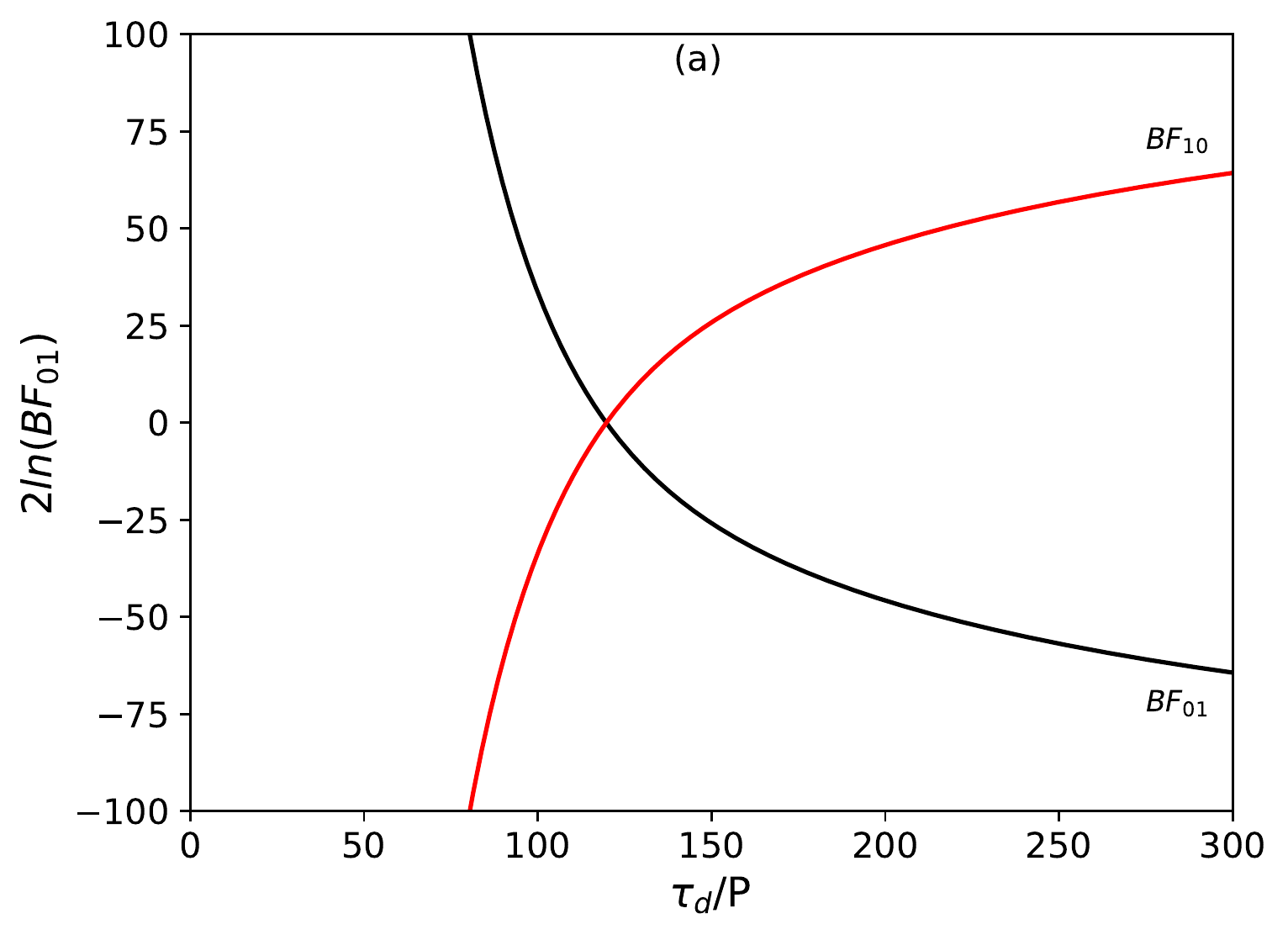} 
	\includegraphics[scale=0.35]{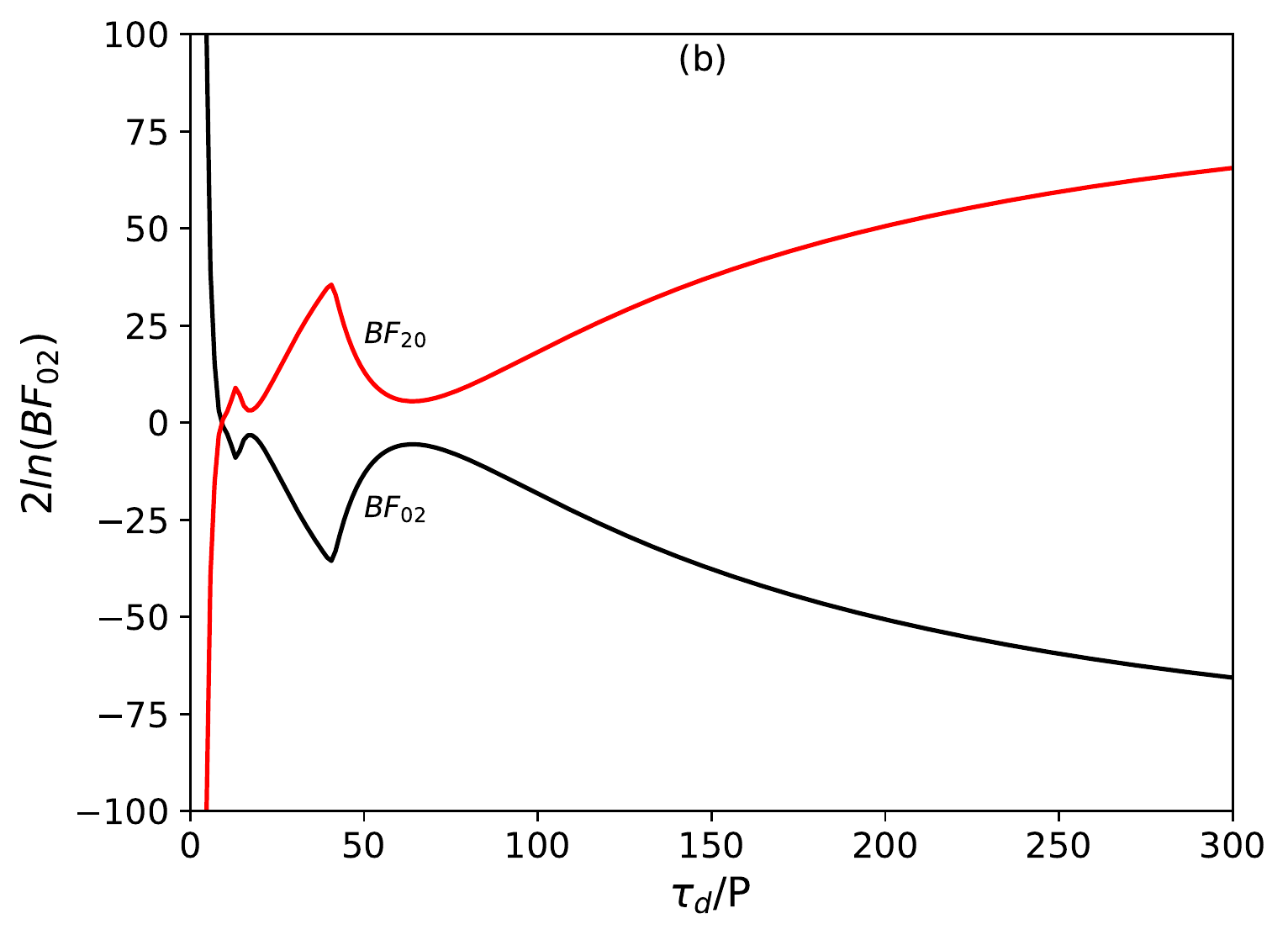}
	\includegraphics[scale=0.35]{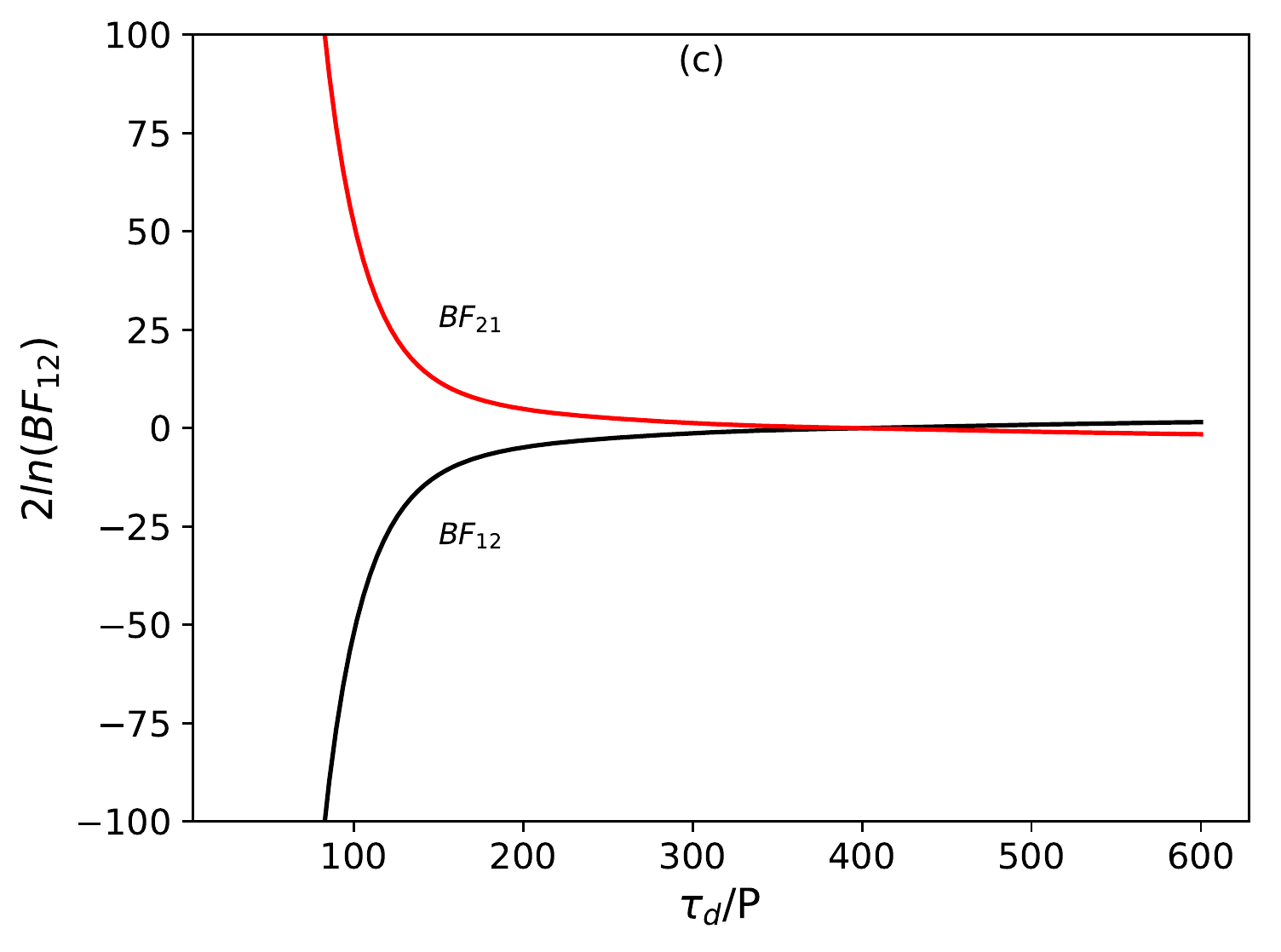}
	
	\caption{Bayes factors as a function of the damping ratio ($\tau_{\rm d}/P$) with an associated uncertainty of 10\%. Subscripts 0, 1, and 2 correspond to resonant absorption in the Alfvén continuum, resonant absorption to the slow continuum, and Cowling's diffusion respectively. \label{f12}}
\end{figure}
\begin{figure*}[!h]
	\centering
	\includegraphics[scale=0.46]{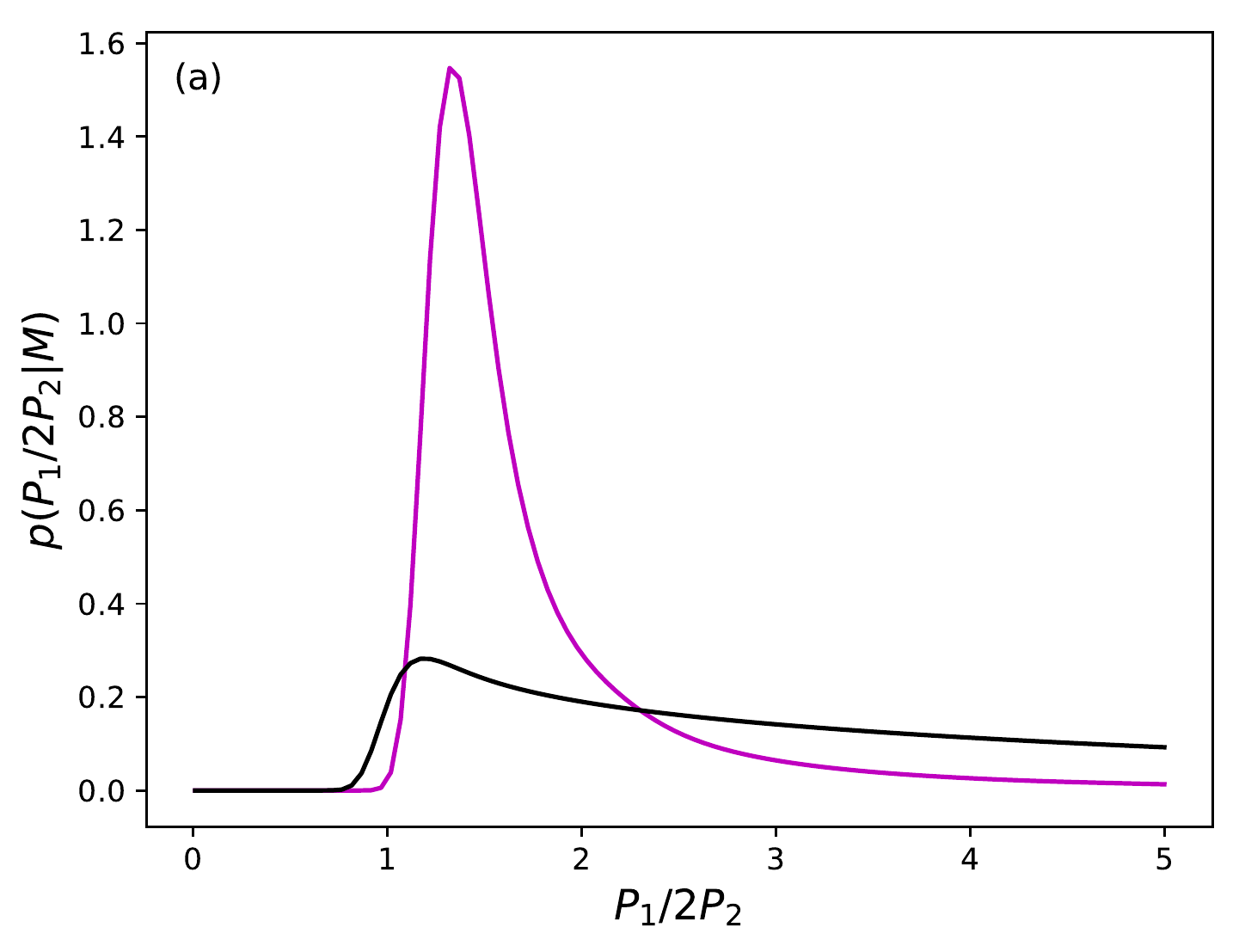}
	\includegraphics[scale=0.46]{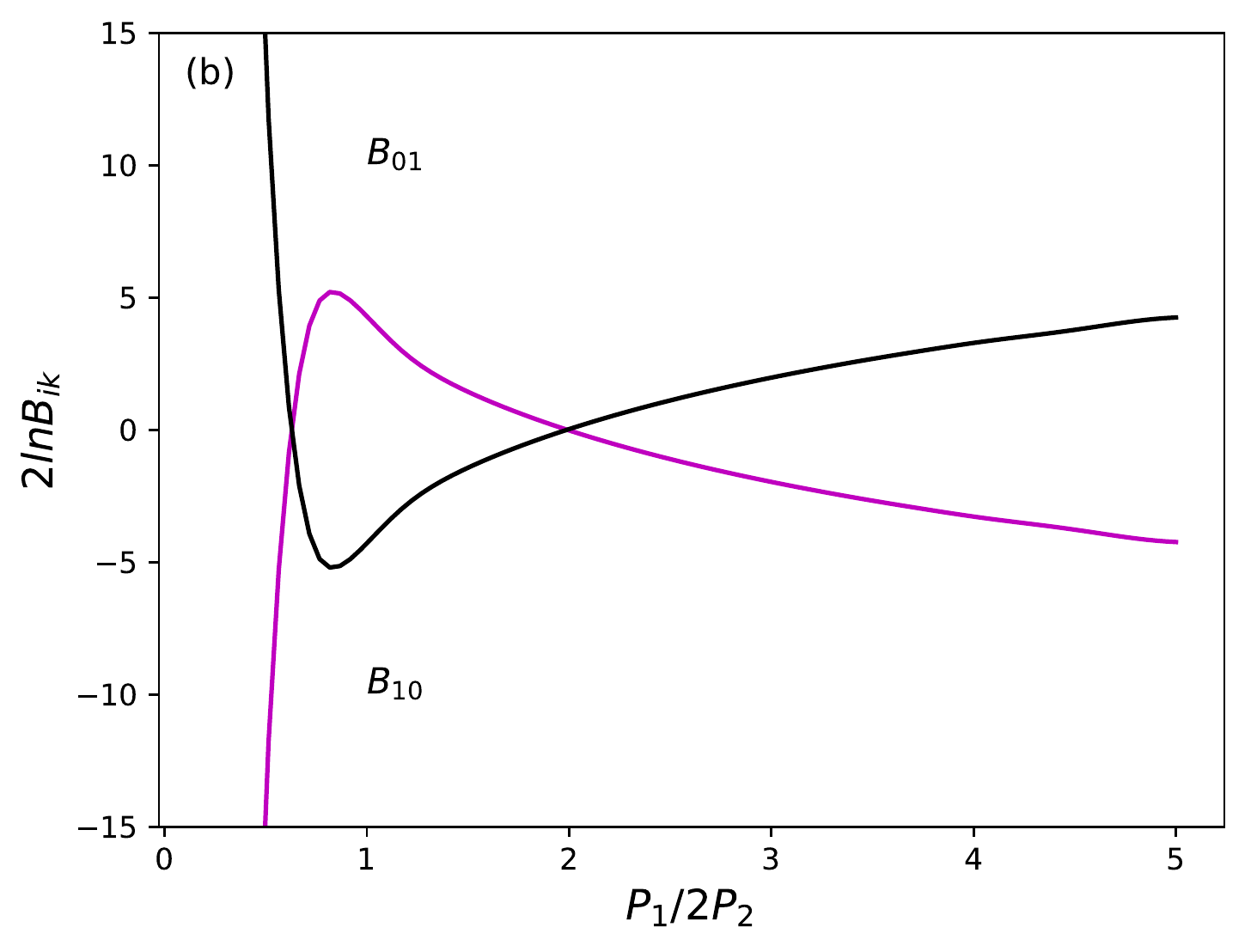}
	\caption{(a) Marginal likelihoods for long and short thread limits. $P_1/2P_2\in(0,5]$ with associated uncertainties of 10\%. (b) Bayes factors as a function of the ratio between the fundamental kink mode period and twice it first overtone period. An uncertainty of 10\% has been assumed. Subscripts 0 and 1 correspond to long and short thread limits respectively.\label{f13}}
\end{figure*}

The marginal likelihood only gives information on how the plausibility of a model is distributed over possible values of observable data. To assess the relative plausibility between models, Bayes factors using Eq.~(\ref{eq4}) must be computed. This is done in a one-to one comparison between the three mechanisms. Then, the obtained Bayes factor values are associated with levels of evidence, according to Table \ref{tab1}. Figure~\ref{f12} shows  Bayes factors, $B_{\rm ij}$, as a function of the observable damping ratio, with the subscripts ${\rm ij} \in [0,1,2]$ corresponding to resonant absorption in the Alfvén continuum, resonant absorption in the slow continuum, and Cowling's diffusion respectively. 
Hence, when confronted to the other alternatives, resonant absorption in the Alfvén continuum is the only mechanism for which Bayes factors showing strong evidence is obtained in the region of typically observed damping ratios of the order of units.

In the comparison between resonant absorption in the Alfvén continuum and resonant absorption in the slow continuum (Fig.~\ref{f12}a), we find strong evidence in favour of the first model for damping ratios up to 125, with Bayes factor values reaching up to 100 (the figure was cut in the vertical direction for clarity). In the rest of the interval, resonant absorption in the slow continuum dominates with very strong evidence.
In the comparison between resonant absorption in the Alfvén continuum and Cowling's diffusion (Fig.~\ref{f12}b), we find strong evidence in favour of resonant absorption for the lowest values of damping ratio ($\tau_{\rm d}/P<9$), intermediate values do not show enough evidence to support any of the two models and the rest of values ($\tau_{\rm d}/P>13$) are better explained by Cowling's diffusion. Finally, in the comparison between resonant absorption in the slow continuum and Cowling's diffusion (Fig.~\ref{f12}c), we find very strong evidence in favour of Cowling's diffusion for all damping ratios below $\tau_{\rm d}/P\sim 200$. For the remaining damping ratio values, the evidence is not large enough to support any of these two models.

\citet{Soler2009d} already found that the damping time due to resonant absorption in the Alfvén continuum is much shorter than the one corresponding to resonant absorption in the slow continuum. In a situation with both, the former would damp the oscillations much earlier. Our analysis is based on reasoning about plausibilities in the Bayesian context. In this case, the computation of marginal likelihoods has to be done separately and gives a measure of how many times different combination of parameters would lead to a given observable. 

\subsubsection{Short and long thread limits}

In Section~\ref{sec:periods} thread lengths and densities were inferred under the short and long thread approximations for the period ratio. As the length of the full magnetic tube is difficult to estimate, we wish to ascertain how plausible each approximation is for a given observationally estimated period ratio. To do so, we consider Eqs.~(\ref{eq16}) and (\ref{eq17}) and compute the marginal likelihood associated to each approximation. We also compare them through the use of Bayes factors. The ranges of plausible values for the parameters, $\boldsymbol{\theta}=\{L_{\rm p}/L, \rho_{\rm p}/\rho_{\rm c}\}$, are the same as in inference analysis. Uniform priors have been assumed for these parameters and values of the period ratio, $P_1/2P_2$, from 0 to 5 are taken with an uncertainty of 10\%.

Figure~\ref{f13}a shows the obtained marginal likelihoods. We can appreciate that values of $P_1/2P_2$ smaller than 1 are very unlikely under the considered models. For the long thread approximation, the marginal likelihood peaks around period ratios in between 1 and 2.5 approximately, with probabilities being 3 times larger that for the short thread limit. For larger values of the period ratio, the short thread approximation seems to be more probable.

Regarding Bayes factors, the results are shown in Fig.~\ref{f13}b. Very strong evidence in favour of the short thread model is obtained when $P_1/2P_2$ values are below 1, positive evidence in favour of the long thread model for intermediate values of the period ratio, and positive evidence for the short thread limit for even larger period ratio values.

Therefore, in partially filled tubes, the long thread approximation seem to be capable of better explaining values of the period ratio in between 1 and 2.5, while the short thread approximation has a larger evidence for period ratio values below and above that range. This model comparison analysis enables us to quantify the goodness of the inference results in section \ref{sec:periods} under a given period ratio approximation for a given period ratio measurement, with its corresponding observational error.

\section{Summary and conclusions}\label{sec:conclusions}

We applied Bayesian analysis techniques to infer physical properties of prominence threads and to quantify the plausibility of alternative models using prominence seismology. 

We first considered fully and partially filled density tubes to model prominence threads and infer their equilibrium characteristics. Our results indicate that the magnetic field strength of prominence threads can be properly inferred and magnitudes smaller than 20 G are obtained for a totally filled tube and even smaller for a partially filled tube, with a rather large variability depending on the particular thread that is considered in a quiescent prominence. The damping model parameters of three damping mechanisms: resonant absorption in the Alfvén continuum, resonant absorption in the slow continuum, and Cowling's diffusion can also be inferred giving information on the transverse inhomogeneity length scale, the wavenumber, plasma-$\beta$, and Cowling's diffusion coefficient. Observations of period ratios between the fundamental kink mode and the first longitudinal overtone enable to infer the lengths of prominence threads in relation to the total length of the flux tube. Different results are obtained depending on whether the long or short thread approximations are considered. The analysis also leads to the conclusion that thread densities larger than 200 have a very low probability. When flows are included in the modelling, the total length of the flux tube can be inferred leading to values in between 20-40 Mm in a particular application to a prominence observed with Hinode/SOT. These values are lower than those  expected for an active region prominence. In a more general case, the total length of the flux tube seems to depend on the flow velocity measurements more than on the observed length of the thread.

Two applications of Bayesian model comparison were presented. Our comparison between three alternative damping mechanisms shows that resonant absorption in the Alfvén continuum is the most plausible mechanism for explaining current observations with very short damping times. The comparison between two analytical approximations for the period ratio of thread oscillations indicates that values of the period ratio around 1 are better explained by the long thread model, while period ratio values below 1 and above 2.5 are better explained by the short thread model. 

We believe the application of Bayesian techniques to prominence seismology is advantageous, because it enables us to constrain what can be plausibly said about difficult to measure physical parameters and to compare the performance of alternative physical models in view of observed data with their uncertainties.

\begin{acknowledgements}
We acknowledge financial support from the Spanish Ministry of Economy and Competitiveness (MINECO) through projects AYA2014-55456-P (Bayesian Analysis of the Solar Corona), AYA2014-60476-P (Solar Magnetometry in the Era of Large Telescopes), and from FEDER funds. M.M-S. acknowledges financial support through a Severo Ochoa FPI Fellowship under the project SEV-2011-0187-03. 
\end{acknowledgements}

\bibliographystyle{aa} 
\bibliography{biblio} 

\end{document}